\documentclass[12pt,letterpaper,twoside]{article}
\usepackage{color,graphicx,natbib,lscape,here,geometry,setspace,multirow,enumitem,colortbl}
\usepackage[dvipsnames]{xcolor}
\usepackage{graphicx}
\usepackage{authblk}
\usepackage{silence}
\usepackage{rotating}
\usepackage{multirow}

\usepackage{booktabs}
\usepackage{color}
\usepackage{setspace}
\usepackage{algorithm2e}
\usepackage{enumitem}
\usepackage{tikz}
\usetikzlibrary{shapes}
\usepackage{standalone}

\usepackage{newtxtext,newtxmath}
\newcommand*{\scfont}{\fontfamily{ptm}\selectfont}

\definecolor{nblue}{HTML}{000660}
\definecolor{yellowXLS}{RGB}{255,255,0}
\definecolor{dorangeXLS}{RGB}{237,125,49}
\definecolor{goldXLS}{RGB}{255,192,0}
\definecolor{lorangeXLS}{RGB}{248,203,173}
\definecolor{greyXLS}{RGB}{174,170,170}
\definecolor{brownXLS}{RGB}{191,143,0}
\definecolor{redXLS}{RGB}{255,80,80}
\definecolor{greenXLS}{RGB}{146,208,80}
\definecolor{purpleXLS}{RGB}{180,130,218}
\definecolor{blueXLS}{RGB}{91,155,213}
\usepackage[colorlinks=true,urlcolor=nblue,linkcolor=nblue,citecolor=nblue]{hyperref}
\usepackage{bigstrut}

\usepackage{tabularx}
\usepackage{threeparttable}
\usepackage{dcolumn}

\usepackage{etoolbox} 
\makeatletter
\patchcmd{\BR@backref}{\newblock}{\newblock[}{}{}
\patchcmd{\BR@backref}{\par}{]\par}{}{}
\makeatother

\usepackage{pdflscape}
\usepackage{afterpage}

\usepackage{longtable}
\usepackage{array}
\newcolumntype{C}[1]{>{\centering\arraybackslash}p{#1}}

\usepackage{comment}
\usepackage{mathtools}

\usepackage[hang,flushmargin]{footmisc} 
\usepackage[english]{babel}

\pdfoutput=1
 \usepackage{float}
 \restylefloat{table}

\newcommand*\xbar[1]{%
   \hbox{%
     \vbox{%
       \hrule height 0.7 pt 
       \kern 0.2ex
       \hbox{%
         \kern-0.05em
         \ensuremath{#1}%
         \kern-0.05em
       }%
     }%
   }%
} 

\newcommand*\xAbar[1]{%
   \hbox{%
     \vbox{%
       \hrule height 0.7 pt 
       \kern 0.16ex
       \hbox{%
         \kern-0.05em
         \ensuremath{#1}%
         \kern-0.05em
       }%
     }%
   }%
} 

\newcommand*\uxbar[1]{%
   \hbox{%
     \vbox{%
       \hrule height 0.7 pt 
       \kern -1.8ex
       \hbox{%
         \kern-0.05em
         \ensuremath{#1}%
         \kern-0.05em
       }%
     }%
   }%
} 

\usepackage[title,titletoc]{appendix}
\makeatletter

\renewenvironment{appendices}{%
    \begin{oldappendices}%
    \renewcommand{\thefigure}{\ifnum \c@section>\z@ \thesection.\fi\@arabic\c@figure}%
    \@addtoreset{figure}{section}%
    \renewcommand{\thetable}{\ifnum \c@section>\z@ \thesection.\fi\@arabic\c@table}%
    \@addtoreset{table}{section}}{%
    \end{oldappendices}%
}\makeatother

\usepackage{titlesec} 
\titleformat{\section}[block]{\large}{\thesection. }{0em}{\MakeUppercase} 
\titleformat{\subsection}[block]{\large}{\thesubsection. }{0em}{\itshape} 
\titleformat{\subsubsection}[block]{\large}{}{0em}{\itshape} 

\let\natbibcitet\citet
\renewcommand\citet{\bibpunct{(}{)}{,}{a}{,}{,}\natbibcitet}
\let\natbibcitep\citep
\renewcommand\citep{\bibpunct{(}{)}{;}{a}{,}{;}\natbibcitep}
\newcommand{\bi}{\begin{itemize}}
\newcommand{\ei}{\end{itemize}}
\newcommand{\be}{\begin{equation}}
\newcommand{\ee}{\end{equation}}
\defcitealias{ieo14}{IEO, 2014}

\long\def\symbolfootnote[#1]#2{\begingroup%
\def\thefootnote{\fnsymbol{footnote}}\footnote[#1]{#2}\endgroup}

\makeatletter
\def\ubar#1{\underline{\sbox\tw@{$#1$}\dp\tw@\z@\box\tw@}}
\def\obar#1{\overline{\sbox\tw@{$#1$}\dp\tw@\z@\box\tw@}}
\makeatother

\usepackage{bm}
\usepackage{caption}
\usepackage{subcaption}

\makeatletter\let\p@subfigure\thefigure\makeatother

\floatstyle{plaintop}
\restylefloat{table}

\usepackage{fancyhdr} 
\usepackage{cleveref}
\crefname{chapter}{Chapter}{Chapters}
\crefname{section}{Section}{Sections}
\crefname{subsection}{Section}{Sections}
\crefname{subsubsection}{Section}{Sections}
\crefname{figure}{Figure}{Figures}
\crefname{table}{Table}{Tables}
\crefname{equation}{Equation}{Equations}
\crefname{appendix}{Appendix}{Appendices}
\crefname{appendices}{Appendix}{Appendices}

\crefname{appsec}{Appendix}{Appendices}

\usepackage{catoptions}
\makeatletter

\def\Autoref#1{%
  \begingroup
  \edef\reserved@a{\cpttrimspaces{#1}}%
  \ifcsndefTF{r@#1}{%
    \xaftercsname{\expandafter\testreftype\@fourthoffive}
      {r@\reserved@a}.\\{#1}%
  }{%
    \ref{#1}%
  }%
  \endgroup
}
\def\testreftype#1.#2\\#3{%
  \ifcsndefTF{#1autorefname}{%
    \def\reserved@a##1##2\@nil{%
      \uppercase{\def\ref@name{##1}}%
      \csn@edef{#1autorefname}{\ref@name##2}%
      \autoref{#3}%
    }%
    \reserved@a#1\@nil
  }{%
    \autoref{#3}%
  }%
}
\makeatother

\newcolumntype{d}[1]{D{.}{.}{#1}}

\title{
{{Capital Flows and the Stabilizing Role of Macroprudential Policies in CESEE}
\thanks{E-mail: \href{mailto:markus.eller@oenb.at}{markus.eller@oenb.at}, \href{mailto:niko.hauzenberger@sbg.ac.at}{niko.hauzenberger@sbg.ac.at}, \href{mailto:florian.huber@sbg.ac.at}{florian.huber@sbg.ac.at},    \href{mailto:helene.schuberth@oenb.at}{helene.schuberth@oenb.at} and \href{mailto:lukas.vashold@s.wu.ac.at}{lukas.vashold@s.wu.ac.at}. Opinions expressed by the authors do not necessarily reflect the official viewpoint of the Oesterreichische Nationalbank or of the Eurosystem. The authors would like to thank Zoltan Walko and Michael W\"{o}gerer for excellent research assistance and Alina Boba{\c s}u, Reiner Martin, Claudia Maurini, Francesco Mazzaferro, Esther Segalla, Ursula Vogel as well as participants of the following events for helpful comments and valuable suggestions: the $4^{th}$ \href{https://ceseephd.net/ceseenet-annual-phd-workshops/}{CESEEnet} research workshop (March 2019 in Mauerbach), the 2019 Annual Meeting of the Central Bank Research Association (CEBRA, July 2019 in New York), the OeNB's 2019 Conference on European Economic Integration (November 2019 in Vienna), the $17^{th}$ ESCB Emerging Markets Workshop (December 2019 at the OeNB), the 2020 Annual Meeting of the Austrian Economic Association (NOeG, February 2020 in Vienna) and seminars held at the OeNB, the ECB, the European Systemic Risk Board (ESRB), the Joint Vienna Institute (JVI), the University of Salzburg and the Austrian Institute of Economic Research (WIFO).}
}}
\date{}
\usepackage{authblk}

\author[1]{Markus Eller}
\author[2]{Niko Hauzenberger}
\author[2]{Florian Huber}
\author[1]{Helene Schuberth}
\author[3]{Lukas Vashold}
\affil[1]{Oesterreichische Nationalbank (OeNB)}
\affil[2]{Salzburg Centre of European Union Studies (SCEUS), University of Salzburg}
\affil[3]{Vienna University of Economics and Business}


\def\equationautorefname~#1\null{%
  Eq.~(#1)\null
}
\def\equationautorefname~#1\null{
Eq.~(#1)\null
}

\begin{document}

\maketitle

\vspace*{-1cm}


\begin{abstract}
\noindent In line with the recent policy discussion on the use of macroprudential measures to respond to  cross-border risks arising from capital flows, this paper tries to quantify to what extent macroprudential policies (MPPs) have been able to stabilize capital flows in Central, Eastern and Southeastern Europe (CESEE) -- a region that experienced a substantial boom-bust cycle in capital flows amid the global financial crisis and where policymakers had been quite active in adopting MPPs already before that crisis. To study the dynamic responses of capital flows to MPP shocks, we propose a novel regime-switching factor-augmented vector autoregressive (FAVAR) model. It allows to capture potential structural breaks in the policy regime and to control -- besides domestic macroeconomic quantities -- for the impact of global factors such as the global financial cycle. Feeding into this model a novel intensity-adjusted macroprudential policy index, we find that tighter MPPs may be effective in containing domestic private sector credit growth and the volumes of gross capital inflows in a majority of the countries analyzed. However, they do not seem to generally shield CESEE countries from capital flow volatility. 
\end{abstract}

\medskip
\begin{tabular}{p{0.2\hsize}p{0.65\hsize}} 
\textbf{Keywords:} & Capital flows, macroprudential policy, global factors, regime-switching FAVAR, CESEE\\
\end{tabular}

\smallskip

\begin{tabular}{p{0.2\hsize}p{0.4\hsize}}
\textbf{JEL Codes:} & C38, E61, F44, G28. \\
\end{tabular}
\vspace{0.25cm}
\begin{center}
\date{\small{\today}}
\end{center}
 
\bigskip

\renewcommand{\thepage}{\arabic{page}}
\setcounter{page}{1}


\section{Introduction}
\label{sec:intro}
The aim of macroprudential policy (MPP) is to avoid macroeconomic costs associated with financial instability, since there is no direct channel for monetary policy to sufficiently guarantee financial stability \citep[][]{galati2013macroprudential, svensson2018monetary}. Policymakers, however, face several challenges when conducting macroprudential policies. One major issue is the lack of a clear and obvious measure for financial stability \citep{svensson2018monetary}. Moreover, new challenges have emerged, due to highly interconnected international financial markets,  involving large swings in international capital (financial) flows and also cross-border spillovers of macroprudential policy measures. 

Volatile capital flows are often seen as a major source for boom-bust cycles in credit or asset prices, eventually impacting financial sector stability. The volatility of capital flows, in turn, is apparently strongly affected by global ``push'' factors, such as the global financial cycle \citep[see, for instance,][]{calvoleidermanreinhart1996, fratzscher2012, rey2015dilemma, lepers-mercado-2020, eller-huber-schuberth-2020}. As a result, the question arises how effective macroprudential measures actually can be in shielding countries from globally determined capital flow volatility. The related policy debate is already quite advanced, discussing the capability of MPPs in increasing the resilience to volatile capital flows and in complementing traditional capital flow management measures \citep[e.g.][]{beirne2014, imf-pp-2016, imf-pp-2017, lepers-mehigan-2019}. 

Quantifying the empirical effects of macroprudential policies is a quickly emerging field -- not least thanks to newly available databases that capture the implementation of specific MPP measures across the globe. There is already a large literature on the efficacy of MPP measures to tame domestic credit cycles and some papers establish a link to capital flow dynamics \citep[e.g.][]{ostry2012tools,forbes2015capital, aizenman2017financial, beirne2017,fendouglu2017credit, igan2017capital}. A small, but growing, strand of the literature addresses the efficacy of MPPs to stabilize domestic macroeconomic quantities \citep[e.g.][]{kim2018effects, richter-etal-2018}.

Nevertheless, there are only a few (working) papers that have already studied the \emph{direct} response of capital flows to MPP measures despite the intense policy debate. \citet{aysan2014mpp} find that cross-border capital flows to Turkey were less sensitive to global factors after the implementation of MPPs in late 2010. \citet{ceruttizhou2018} study, for a huge panel of countries over the period 2006--2015, the joint impact of macroprudential and capital control measures on cross-border banking flows: Tighter MPPs in lender countries apparently reduce direct cross-border banking outflows but are associated with larger outflows via local affiliates. Tighter MPPs in borrower countries, on the other hand, are associated with larger direct cross-border banking inflows, likely due to circumvention motives. In a similar vein, \citet{frost-etal-2020} study a large panel of countries for the time period of 2000--2017 and find that the activation of foreign exchange (FX)-based MPPs reduces capital inflow volumes by nearly 5\,\% of GDP and is linked to a lower probability of banking crisis and capital flow surges in the following three years. \citet{ahnert-etal-2018} show for a sample of 48 countries that spans the period from 1996 to 2014 that macroprudential FX regulation of banks is effective in reducing banks' FX borrowing but also has the unintended consequence of simultaneously causing firms to increase FX bond issuance and thus shifting FX exposure to other sectors of the economy. 


This paper tries to contribute to the existing literature along the following dimensions. 
\textit{First}, in terms of regional focus, we investigate the countries from Central, Eastern, and Southeastern Europe (CESEE). As small open economies, they show considerable external vulnerabilities -- among others due to a large share of public and private sector debt being denominated in foreign currency -- and are therefore susceptible to global risk fluctuations that often result in sudden shifts in international capital flows. On the other hand, several CESEE countries had been quite active in implementing MPPs already before the global financial crisis (GFC), and thus for a much longer period than countries in western Europe, mostly to rein in extraordinarily strong credit growth at the time. Macroprudential measures are expected to have a measurable effect on cross-border banking flows, which are of particular importance in the CESEE countries given the prominent role of foreign parent banks. 
\textit{Second}, in terms of econometric methodology, we propose a regime-switching factor-augmented vector autoregression (FAVAR) framework, while most of the previous cross-country studies have relied on simple fixed-effects panel regressions. Our model allows studying country-specific capital flow responses to MPP shocks, capturing the dynamics in a closed economy and accounting at the same time for global (exogenous) factors -- such as the global financial cycle -- in a parsimonious framework. Moreover, we allow for nonlinearities in the form of regime switches to capture potential macroprudential policy shifts (e.g. in the wake of high- and low-interest rate episodes). The FAVAR model effectively controls for a large number of external indicators, summarized in few factors, and is considered a valid method to conduct structural analysis \citep{bernanke2005measuring}.
\textit{Third}, in terms of data, we utilize a novel intensity-adjusted macroprudential policy index \citep[MPPI,][]{eller-etal-taxonomy}. In contrast to most of the literature that captures only the occurrence of MPPs using rather simple indices, this index allows us to track not only if, but also to what extent a measure was implemented.\footnote{To give an example, it should make a difference for any impact assessment of macroprudential policy tightening if we treated a lowering of the maximum loan-to-value (LTV) ratio from 100\,\% to 60\,\% in a different way than a reduction from 100\,\% to only 90\,\%. Many existing investigations would just use a dummy approach and would treat both cases identically.} Next to the intensity adjustment, this index covers a comparatively long time span (of more than 20 years), captures a large variety of instruments and differentiates between the announcement and implementation of measures.

The remainder of the paper is structured as follows: Section~\ref{sec:econ} describes the details of the nonlinear FAVAR framework as well as the prior specification. Section~\ref{sec:data} provides more details on the macroprudential policy index used and describes the macroeconomic data. Section~\ref{sec:ident} explains how we identify an MPP shock. Section~\ref{sec:transmission} then gives an overview of the direct transmission channels through which macroprudential measures are expected to affect capital flows. Section~\ref{sec:results} provides an overview of the related impulse-response results and section~\ref{sec:con} concludes.

\section{Econometric framework}
\label{sec:econ}
In this section we propose a novel factor-augmented vector autoregressive (FAVAR) framework \citep{bernanke2005measuring} with regime-switching in order to assess the effects of macroprudential policy actions for several CESEE countries over time, while controlling for the impact of co-movement in international financial series. After describing key model features, we discuss prior specification and implementation.

\subsection{The nonlinear factor-augmented VAR model}\label{sec: econometrics}
Our approach is based on modeling a set of macroeconomic and financial quantities specific to country $i=1,\dots,N$ while capturing international movements in financial quantities. For country $i$, we assume that a set of $m$ endogenous variables $\bm y_{it}$, including the MPPI, domestic macroeconomic and financial variables as well as a capital flow series $c_{it}$ and its volatility proxy $v_{it}$, depend on their own lags plus lags of a global financial factor (extracted from international financial variables, see section \ref{sec:data}, and represented by a $q$-dimensional vector $\bm F_{it}$). 
Global co-movement in financial sector variables might have triggered similar macroprudential policy decisions in the sample. The inclusion of the global financial factor thus allows to control for global (unidirectional) spillovers.\footnote{As highlighted in several works of the ESRB \citep[e.g.][]{ESRB-2020}, cross-border spillovers and leakages of domestic macroprudential measures provide a rationale for stronger cross-border coordination of macroprudential policies (including reciprocation of measures). Let us assume that country $i$ implements a macroprudential tightening; if other important partner countries responded reciprocally, this could also have an impact on capital flows to country $i$. As a caveat, though, we cannot account for such bilateral spillovers, since we are estimating country-specific VARs. A global VAR or a panel VAR could be a solution to account for bilateral cross-border linkages; at the same time, it would be quite a challenge in these settings to properly identify the policy shock and to account for different effects over time.} 
Defining $\bm{x}_{it} = (\bm F_{it}', \bm y_{it}')'$ allows us to establish a relationship between the observed quantities (a set of international macroeconomic and financial quantities stored in an $S$-dimensional vector $\bm Z_{it}$) and the observed and unobserved factors in $\bm x_{it}$,


\begin{equation}
\begin{pmatrix}
\bm Z_{it} \\ \bm y_{it}
\end{pmatrix} = \begin{pmatrix}
\bm \Lambda_{i S_t} & \bm 0 \\
\bm 0 & \bm I
\end{pmatrix} 
\begin{pmatrix}
\bm F_{it} \\ \bm y_{it}
\end{pmatrix}
+ \begin{pmatrix}
\bm \eta_{it}\\
\bm 0
\end{pmatrix}. \label{eq:measurement}
\end{equation}
Here, $\bm {\Lambda}_{i S_t}$ denotes a $(S \times q)$-dimensional matrix of regime-specific factor loadings with $S \gg q$, and $S_{it}$ is an endogenous regime indicator.
We assume that $S_{it} \in \{0,1\}$ follows an endogenous Markov switching process that is driven by the country-specific short-term interest rate $i_{it}$, with transition probabilities discussed in section \ref{sec: transprobs}. This model feature will allow us to study later in the impulse-response analysis whether responses to MPP shocks differ over time, distinguishing between high- and low-interest rate episodes. Finally, $\bm \eta_{it}$ represents an $S$-dimensional vector of measurement errors that follow a Gaussian distribution with mean zero and diagonal variance-covariance matrix $\bm \Sigma_{i}=\text{diag}(\sigma_1^2, \dots, \sigma_S^2)$.


Equation (\ref{eq:measurement}) constitutes the measurement equation that relates observed to latent quantities. A few features are worth discussing. 
\textit{First}, we need an identifying assumption on  $\bm \Lambda_{iS_t}$. In what follows, we assume that the upper $q \times q$ block of $\bm \Lambda_{iS_t}$ is set to an identity matrix $\bm I_q$. Estimating the latent factors by means of principal components (PCs) then corresponds to extracting PCs from the corresponding set of observed quantities. 
\textit{Second}, we assume that the factor loadings are regime-specific. This implies that the sensitivity of elements in $\bm Z_{it}$ with respect to movements in $\bm F_{it}$ is time-varying and changes across two economic regimes. \textit{Third}, any comovement in $\bm Z_{it}$ stems exclusively from the latent factors $\bm F_{it}$.

The latent states and observed quantities in $\bm x_{it}$ are then assumed to follow a regime-switching VAR model of order $P$,
\begin{equation}
\begin{aligned}
\bm x_{it}
=& \bm a_{0,iS_t} + \sum_{p = 1}^{P} 
\bm{A}_{p,iS_t} \bm{x}_{it-p} 
+ \bm \epsilon_{it}, \quad \text{with} \quad \bm \epsilon_{it} \sim \mathcal{N} \left( \bm 0, \bm \Omega_{iS_t} \right),
\end{aligned}
\end{equation}
whereby $\bm a_{0,iS_t}$ denotes a $K$-dimensional vector of state-specific intercepts ($K = m + q$), $\bm{A}_{p,i S_t}~(p=1,\dots,P)$ is a $K \times K$-dimensional matrix of state-specific coefficients, while $\bm \epsilon_{it}$ is a set of Gaussian shocks with zero mean and regime-specific variance-covariance matrix $\bm \Omega_{iS_t}$.

Up to this point, we remained silent on how to obtain the volatility estimates. In what follows, we simply assume that $v_{it}$ is obtained by first estimating autoregressive models of order $r$ on the corresponding capital flow series $c_{it}$,
\begin{equation}\label{eq:ar-p}
\begin{aligned}
c_{it} = \sum_{m=1}^r \rho_m c_{it-m} + e_{it}
\end{aligned}
\end{equation}
with $e_{it}$ being a zero-mean  Gaussian shock with time-varying variance $\exp(v_{it})$. The (log) variance then follows an AR(1) process,
\begin{equation}\label{eq:sv}
\begin{aligned}
v_{it} = \mu_{i} + \phi_{i} (v_{it-1} - \mu_{i}) + \varsigma_{it},
\end{aligned}
\end{equation}
with $\mu_{i}$ denoting the unconditional mean of the log-variance, $\phi_{i}$ reflecting the persistence parameter and $\varsigma_{it}$ being a Gaussian shock to the log-volatility with zero mean and variance $\sigma_{iv}^2$. 
After obtaining point estimates of these measures (i.e. the posterior mean), we include the time-varying log-variances $v_{it}$ as a volatility proxy in our model. Another feasible approach to obtain a volatility measure for each capital flow time series would be to use rolling standard deviations.

\subsection{An endogenous mechanism for the state allocation}\label{sec: transprobs}
So far, we remained silent on how the state allocation $S_{it}$ is obtained. We assume that $S_{it}$ follows a Markov switching process with time-varying transition probabilities.\footnote{For robustness, we also consider $S_{it}$ to be specified deterministically. Here, $S_{it}$, for $i = 1, \dots, N$, equals zero in the period before the GFC (up to $2008$Q$4$) while being equal to unity in its aftermath (starting from $2009$Q$1$).}

The transition probability matrix reads as
\begin{equation}
\boldsymbol{P}_{it} = \begin{pmatrix}
p_{00,it} & p_{01,it} \\
p_{10,it} & p_{11,it}
\end{pmatrix}, \label{eq: transProb}
\end{equation}
with $\sum_{l=0}^{1}p_{kl,it} = 1$ for $k = (0, 1)$. 

The transition probabilities of $S_{it}$,  $\Pr(S_{it} = l | S_{it-1} = k,  \gamma_i,  i_{it})=p_{kl, it}$, are linked to the country-specific short-term interest rate $i_{it}$ through a probit model \citep[see, inter alia,][]{filardo1994business, kim1998business, amisano2013money, huber2018markov}. This captures the notion that MPPs are more likely to be adopted when conventional monetary policy is reaching its limit.  
The probit specification is given by
\begin{align}
p_{kl,it}= \Phi ( c_{0,ki}+\gamma_i' i_{it}),
\label{eq:transprob}
\end{align}
with $c_{0,ki}$ denoting a regime-specific intercept, and $\Phi$ refers to the cumulative distribution function of the standard normal distribution. $ \gamma_i$ measures the sensitivity of the transition probabilities with respect to the country-specific interest rate $i_{it}$.\footnote{With more than two regimes, an alternative would be a logit specification \citep[][]{kaufmann2015k, billio2016interconnections, hauzenberger2020fx}.} Similar to \cite{amisano2013money}, we assume that $\gamma_i$ is fixed across regimes while we allow the intercept to be defined by the previous state $S_{it-1}$. 

It is again worth stressing that the corresponding regime allocation determined by $S_{it}$ is stochastic. The interest rate determines the transitions between states but in a stochastic manner. This implies that if there is substantial evidence in the likelihood that a regime change takes place (not driven by changes in policy rates), our model is capable of detecting this shift.

\subsection{A pooling prior}
Since the VAR coefficients are state-specific and we are interested in the effects of macroprudential policy shocks in different interest rate regimes, we encounter several issues. Most prominently, the length of the time series in the subperiods may be small while the number of parameters to estimate is large. Shrinkage priors in the spirit of \cite{sims1998bayesian} would offer a feasible solution. However, the corresponding impulse-responses would be dominated by the prior, which is centered on a multivariate random walk. 

In the present paper, we follow a different approach and borrow strength from coefficient pooling. More specifically, we stack all regime-specific coefficients in a $k$-dimensional vector $\bm{\beta}_{iS_t}= vec \left\lbrace\left(\bm  a_{0,iS_t}, \bm  A_{1,iS_t}, \dots, \bm  A_{P,iS_t} \right)' \right\rbrace$ (with $k=K (KP + 1)$). In the next step, we assume that the state-specific regression coefficients arise from a common Gaussian distribution given by
\begin{equation}
\bm{\beta}_{iS_t} \sim \mathcal{N}(\bm \beta_{i0}, \bm \Xi_i),
\end{equation}
with $\bm \beta_{i0}$ denoting a common mean vector of dimension $k$ and $\bm \Xi_i=\text{diag}(\xi_{i1}, \dots, \xi_{ik})$ being a $(k \times k)$ variance-covariance matrix  with $\xi_{ij}$ denoting a coefficient-specific variance. The size of $\xi_{ij}$ effectively controls whether a given coefficient should be pushed towards the common mean across both regimes. If $\xi_{ij}$ is large, the $j$th element in $\bm \beta_{i S_t}$, $\beta_{ij, S_t}$, is allowed to differ strongly across regimes, whereas in the opposite case, $\beta_{ij, 0} \approx \beta_{ij,1}$ and thus only little differences across regimes are possible. This specification is closely related to  \cite{verbeke1996linear,allenby1998heterogeneity,fruhwirth2004bayesian} and has been applied in the VAR framework in \cite{huber2018stochastic}.  

Similarly to the VAR coefficients, we also pool across error variance-covariance matrices in the VAR state equation. This is achieved by placing a conjugate hierarchical Wishart prior on $\bm \Omega_{iS_t}^{-1}$ for $S_t = 1, 2$:
\begin{equation}
\begin{aligned}
\bm \Omega_{iS_t}^{-1}\sim & \mathcal{W}(\bm \Psi_i, \psi_i) \quad \text{and} \\
\bm \Psi_i \sim & \mathcal{W}(\bm S_i, s_i),
\end{aligned}
\end{equation}
with hyperparameters $\psi_i = 2.5 + (K - 1)/2$, $s_i = 0.5 + (K - 1)/2$, and $\bm S_i = 100s_i/\psi_i\text{diag}(\hat{\sigma}_{1i}, \dots, \hat{\sigma}_{Ki})$ specified according to \cite{malsinerwalli2016, huber2018stochastic}. $\hat{\sigma}_{ji}$ simply denotes OLS variances of univariate AR(s) processes to consider the original scale of the data.

For the remaining coefficients of the model (i.e. the factor loadings, coefficients of the state equation of the log-volatilities, measurement error variances), our prior setup closely follows the existing literature and is specified to be only weakly informative, if possible. Specifically, we use standard normally distributed priors on the free elements of the factor loadings, inverted Gamma priors that are loosely informative on the measurement error variances, Gaussian priors with mean zero and variance ten on the unconditional mean of the log-volatility process and Gamma priors with scale and shape parameter equal to $1/2$ on $\sigma^2_{iv}$. Finally, we use a Beta distributed prior on $\frac{\phi_i+1}{2} \sim \mathcal{B}(25,5)$ that captures the notion that the log-volatility process is quite persistent.

We estimate the model using a Markov chain Monte Carlo (MCMC) algorithm that consists of standard steps \citep[for details, see][]{huber2018stochastic}.
\section{Data}
\label{sec:data}
Our investigated sample covers the 11 EU member states in CESEE\footnote{Bulgaria (BG), Croatia (HR), the Czech Republic (CZ), Estonia (EE), Hungary (HU), Latvia (LV), Lithuania (LT), Poland (PL), Romania (RO), Slovakia (SK) and Slovenia (SI).} with time series available at quarterly frequency from $2000$Q$1$ until $2018$Q$4$. For each country, we include following groups of variables: (1) a global financial factor, (2) the macroprudential policy index (MPPI), (3) domestic macroeconomic and macrofinancial quantities (real GDP growth, CPI inflation, private sector credit growth, short-term interest rate, equity price growth, (real effective) exchange rate volatility) and (4) the levels and volatilites of gross capital inflows \emph{and} outflows. Thus, the number of endogenous variables is $m = 12$. Notice that we assume that the global financial factor is endogenous with respect to a given economy (but is restricted to react with a one-quarter lag to shocks in the remaining quantities). This assumption might be questionable since each economy we consider might be small relative to the rest of the world. However, if the world factor is truly exogenous we expect our flexible shrinkage prior to capture this and push the corresponding coefficients towards zero.\footnote{In the empirical application we do not report the reactions of the global factor to MPP shocks because, in most cases, they are statistically not different from zero.} Table~\ref{tab:data} gives an overview of the used variables, their respective transformations and the main sources they were obtained from. 

Before providing more details on the applied intensity-adjusted MPPI, let us briefly stress a few issues related to other important system variables. 
\textit{First}, as regards \emph{capital flows} as the main variable of interest, gross inflows and gross outflows enter simultaneously the model to control for potential  dependencies since parts of the inflows are often flowing again out of the economy (inter alia due to special purpose entities or round-trip investment).
We include both their levels (percentage of nominal GDP, cumulative four-quarter moving sums) and their volatilities by running an AR(5)-SV process and extracting the time-varying log-variances of this process, in line with equations (\ref{eq:ar-p}) and (\ref{eq:sv}). 
As functional categories we include either total capital flows (i.e. totaled direct, portfolio and other investment flows) or other investment (OI) flows only. The latter mostly reflect direct foreign lending to resident banks and include more volatile funding sources, such as short-term funding sourced on wholesale markets. Therefore, in terms of macrofinancial risks, this capital flow category is of special interest and the way macroprudential policies can affect it deserves special attention (for a broader discussion, see section \ref{sec:transmission}).

\textit{Second}, to account for the impact of comovement in global financial series and thus to proxy for the impact of the global financial cycle, we extract by means of principal components a \emph{global financial factor} from a set of equity prices, private sector credit growth and private sector deposit growth across 45 countries worldwide \citep[following][]{eller-huber-schuberth-2020}. We extract the first principal component from this panel of time series. This summarizes the bulk of variation and thus provides a rather general measure of the global financial cycle. Other papers \citep[e.g.][]{forbes2015capital, aizenman2017financial, avdjiev2018dollar} include the volatility index (VIX) of the Chicago Board Options Exchange (CBOE) and/or the TED spread as financial risk/uncertainty measures to capture the global financial cycle. However, these proxies are probably too US-centric for our sample and do only cover narrow aspects of the financial cycle, which is why we prefer the specification of a global factor extracted from a broad range of countries and variables. 

\subsection{An intensity-adjusted index for macroprudential policies}
\label{subsec:index}

Most of the literature studying the impact of MPPs has relied on rather simple indices that primarily account for the occurrence, but not for the intensity of measures. Some just rely on binary indicators that signal whether a certain instrument was in place at a given time or not \citep[e.g.][]{reinhardt2015regulatory, cerutti2017use}. Most studies use an index where a tightening measure is coded with $+1$ and a loosening measure with $-1$, while ambiguous ones are not taken into account. By cumulatively summing them up over time, an index of macroprudential tightness can be compiled \citep[e.g.][]{shim2013database, ahnert-etal-2018, alam2019digging}. The simplicity of these indices comes, however, with the drawback of neglecting variations in the intensity (i.e. the strength) of such measures. There are only a few papers that have already (partially) applied an intensity adjustment. Most notably, \citet{vandenbussche2015macroprudential} construct an intensity-adjusted index for 16 CESEE countries to investigate the effects of MPPs on housing prices. \citet{dumivcic2018effectiveness} studied the effectiveness of MPPs in mitigating excessive credit growth and accounted for different intensities of MPPs, using step functions that yield different values when the change of a particular instrument exceeds a certain threshold. \citet{richter-etal-2018} and \citet{alam2019digging} both focus on the effect of loan-to-value (LTV) limits on the broader macroeconomic environment and provide more detailed information regarding the intensity of the use of this instrument.

The index used in this study, introduced and described in detail in \citet{eller-etal-taxonomy}, represents another approach for building an index based on the intensity of MPPs. It includes ``classic'' macroprudential instruments and other requirements motivated by macroprudential objectives (e.g. minimum capital and reserve requirements) and covers the 11 CESEE EU member states from 1997 to 2018 on a quarterly basis. 
By applying a set of different weighting rules for the various incorporated measures \citep[building on][]{vandenbussche2015macroprudential}, a quantitatively meaningful summary statistic of macroprudential activity within a given country is constructed. These rules depend on the nature, particularly on the complexity, of the individual instruments and ensure that  to the extent possible differences in their intensities are reflected in the index. Another innovation of this index is the utilization of information regarding the  timing of MPPs. While most existing studies use only the implementation date of a measure, the index at hand also utilizes information about its announcement date. In many cases, there is a nonnegligible gap between these two dates, especially in the case of capital-based measures. Financial institutions may react to regulatory changes as soon as they are announced \citep[for a similar argumentation see e.g.][]{imf-bis-pp-2016,meuleman2020macroprudential}. Especially in the case of tightening incidents, it can be assumed that banks react instantaneously in order to gradually prepare for meeting new regulations before they become binding (e.g. building up a capital buffer). For loosening instances, this would not be the case, as banks have to fulfill the requirements until the day of actual implementation. For these reasons, we use in our empirical investigation the intensity-adjusted MPPI of \cite{eller-etal-taxonomy} based on announcement dates for tightening and implementation dates for loosening macroprudential measures.

\begin{figure}[ht!]
	\centering
	\includegraphics[width=\linewidth, page=1]{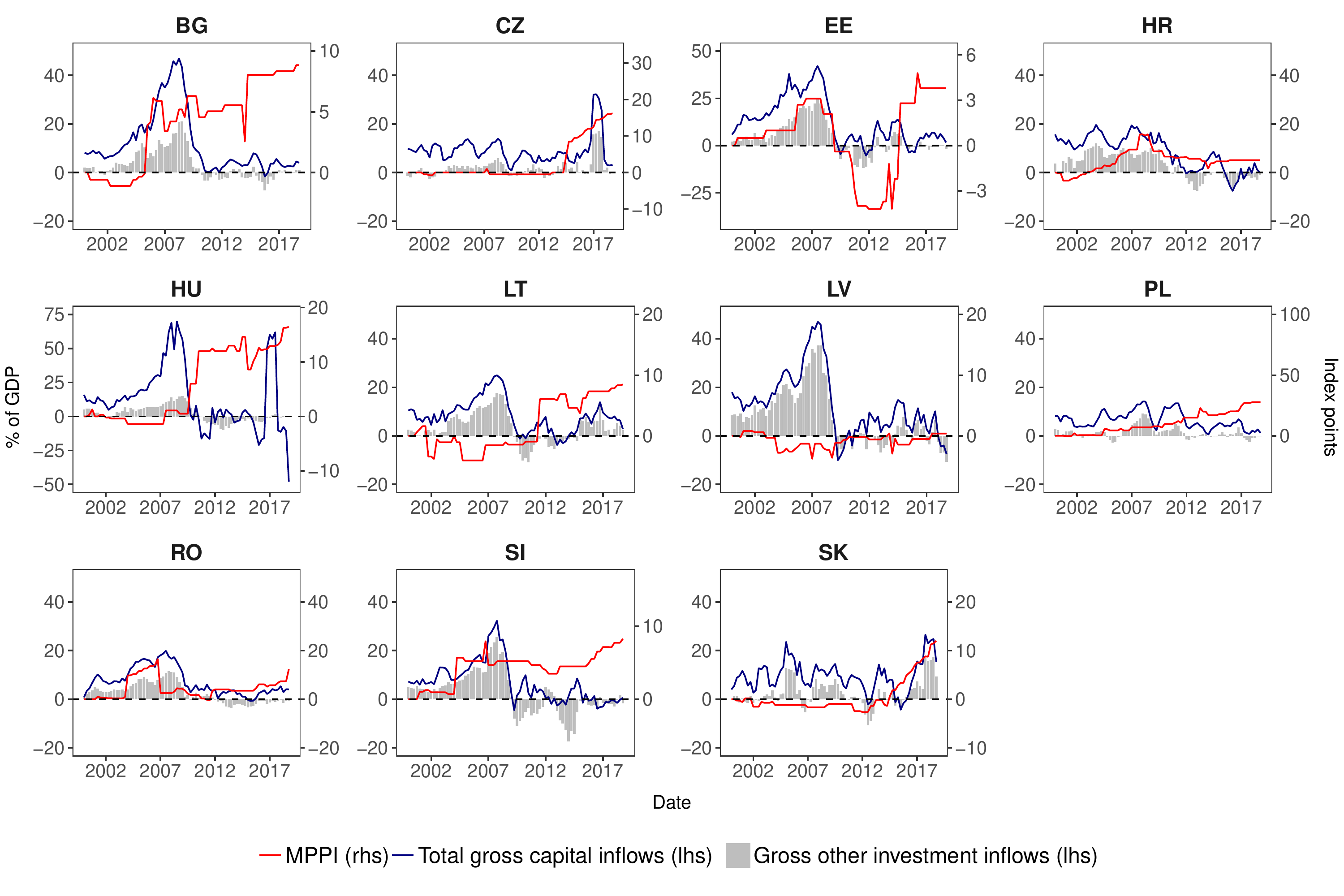}
	\caption{Intensity-adjusted MPPI (red line) using announcement (implementation) dates for tightening (loosening) measures together with gross total capital inflows (blue line) and gross other investment inflows (grey bars), both in \% of GDP, for the time period $2000$Q$1$-$2018$Q$4$. Based on authors' own calculations and data from IMF-IFS. MPPI has been rescaled to start at $0$.}
	\label{fig:mppi_cap}
\end{figure}

Figure~\ref{fig:mppi_cap} shows the MPPI for the countries and time period used in the subsequent estimations together with the capital inflow series. An increase (decrease) in the index signals a net tightening (loosening) in the overall macroprudential environment. It is noteworthy that the patterns across countries are rather heterogeneous. While Bulgaria, Croatia, Poland, Romania and to a certain extent Slovenia already had shown quite some tightening activity before the GFC, a significant pick-up thereof is observable for countries like the Czech Republic, Hungary or Slovakia only after it. In the Baltics, the overall net tightening in the macroprudential environment was not as pronounced, even though activity was apparent throughout the whole period.\footnote{For a closer description of the various country dynamics, also for the period before 2000, the role of different MPP instruments and the detailed construction of the MPPI, we refer to \citet{eller-etal-taxonomy}.} This heterogeneity in the development of MPPs stresses the need to go beyond simple fixed-effects panel regressions and to analyze responses to a macroprudential tightening country-by-country. 
At the same time, we can see in figure~\ref{fig:mppi_cap} that CESEE countries have experienced a substantial boom-bust cycle in capital flows. Before the global financial crisis hit, the CESEE countries attracted sizable gross capital inflows, in a few countries of up to 50\,\% of GDP (Bulgaria, Estonia, Hungary, Latvia). At a global scale, cumulative net capital inflows into the CESEE countries as a percentage of GDP were by far the highest worldwide in the period 2003--2008 and outstripped the flows that poured into East Asia before the Asian crisis hit in the late 1990s \citep{eller2016understanding}. After a sharp  reversal in 2009, capital inflows have recovered somewhat in recent years, but are in most countries still far below the numbers recorded before the GFC. Moreover, gross other investment inflows, or bank flows, made up for a major share of total inflows before the GFC, but in recent years portfolio inflows have gained importance.\footnote{In the case of Hungary, the marked spikes of capital inflows (and outflows, not shown here) at the end of the sample were due to new SPEs that were established at the end of 2016 and left the country again at the end of 2018 (related to the worldwide restructuring of a large multinational pharmaceutic  group).} 
The relation between the MPPI and capital flows is, at first glance, not always clear-cut. Before the GFC, several countries tightened their MPP stance, but capital inflows had still surged. At the same time, we do not know the counterfactual; the surge of capital flows could have been considerably stronger in the absence of the tightened MPP environment. After the GFC, in most, but not all, countries the capital flow reversal coincides with a procyclical tightening of the MPP stance.

\section{Identification}\label{sec:ident}
Theoretical contributions, such as \cite{gerali2010credit}, \cite{akram2014macro} and \cite{angelini2014interaction}, proposing dynamic stochastic general equilibrium (DSGE) models, offer a theoretical justification to identify the transmission channels of an MPP shock in a VAR model. 
In a first step a linear specification of the FAVAR model for the whole model is estimated in order to contrast these results (which can be found in section~\ref{main:single}) with the ones obtained for the specification that allows for a possible regime switch as described in section~\ref{sec:econ}. The latter results (see section~\ref{main:ms}) and their interpretation is the main focus of this study, as they allow for nonlinearities and should thus capture a more accurate picture of the changing effectiveness of macroprudential policies. To inform and update the transition probabilities of the endogenous regime-switching step, we use information about short-term interest rates to identify a high- and a low-interest rate regime. However appendix~\ref{add:detregime} also provides results for a deterministic regime switch set equal to the onset of the GFC.

Moreover, we propose a Cholesky-type identification scheme in table \ref{tab:ident}, assuming that macroprudential policy responds in the period of the shock only to (exogenous) global financial cycle movements, but not to other (faster) variables in the system. These impact restrictions can be  advocated by the legislation process of macroprudential policy, by the long lead time of these measures and by the use of quarterly data. For example, \cite{meeks2017capital} argues that the policy variable does not react immediately to macroeconomic changes and that there is only an indirect transmission through lending channels. Other studies using a similar identification scheme are \cite{kim2017managing} and \cite{kim2018effects}. Alternative identification schemes based on sign restrictions would be feasible. However, using sign restrictions implies that we obtain a \emph{set} of structurally identified impulse-responses.  This, in light of weak identifying assumptions on the impulse-responses, potentially inflates the estimation uncertainty surrounding our impulse-responses, especially in the second part of our sample.

\begin{table}[!htbp]
	\footnotesize
	\begin{tabular}{l|cccccc}
		& Global factor & MPPI & \textbf{Slow macro} & Stir & \textbf{Fast macro-fin} & \textbf{Capital flow} \\
		\hline
		Global factor & $x$ & $0$  &  $\bm 0$     &  $0$ &  $\bm 0$ &  $\bm 0$ \\
		MPPI	& $x$       & $x$  &  $\bm 0$     &  $0$ &  $\bm 0$ &  $\bm 0$  \\
		\textbf{Slow macro}	 		& $\bm{x}$ & $ \bm x$  &  $\bm x$     &  $ \bm 0$ &  $\bm 0$  &  $\bm 0$  \\
		Stir 	& $x$   & $x$  &  $\bm x$     &  $x$ &  $\bm 0$  &  $\bm 0$ \\
		\textbf{Fast macro-fin} & $\bm{x}$ & $\bm{x}$  &  $\bm x$  &  $\bm{x}$ &  $\bm{x}$ &  $\bm{0}$  \\   
		\textbf{Capital flow}   & $\bm x$ & $\bm x$  &  $\bm x$     &  $\bm x$ &  $\bm x$ &  $\bm x$    \\    
	\end{tabular}
	\caption{Identification scheme defining zero impact restrictions. Bold letters indicate a vector of multiple variables. \textbf{Slow macro} covers real GDP growth, CPI inflation and credit growth. \textbf{Stir} captures the short-term interest rate to account for the impact of monetary policies. As \textbf{Fast macro-fin}, we consider equity price growth and the volatility of the REER. \textbf{Capital flow} covers the respective capital in- and outflow series and their volatility proxies.}
	\label{tab:ident}
\end{table}

\section{Transmission channels through which macroprudential policies can affect capital flows}\label{sec:transmission}

In this section, we review the transmission channels from MPP measures on capital flows as discussed in the literature. In principle, macroprudential tools, which do not target the financial account per se, can nevertheless have an impact on capital flows \citep{imf-pp-2017}. Disregarding for a moment the potential different effects of the broad range of macroprudential tools available, the more general channels that link MPP tools with the financial account can be described as follows:

\textit{First}, MPP tools can increase resilience of the financial system and help contain the build-up of systemic financial risks during capital inflow surges and reversals of flows. Excessive capital inflow episodes may be limited by containing the procyclical interplay between asset prices, private credit and non-core bank funding.  By restricting increases in leverage and volatile funding, the resilience with regard to fluctuations in the global financial cycle may be enhanced. \citet{cesa2018international}, for instance, found that countries featuring lower LTV ratios and stricter limits on foreign currency borrowing are less vulnerable to global credit supply shocks. Similarly, \cite{coman2019face} found that tighter LTV limits and reserve requirements appear to be particularly effective measures to shield emerging markets from negative spillover effects of US monetary policy. \citet{bergant-etal-2020} show that tighter levels of macroprudential regulation can considerably dampen the sensitivity of GDP growth in emerging markets with respect to global financial shocks.
Furthermore, during financial stress the incidence of disruptive capital outflows may be reduced. Examples are (countercyclical capital) buffers that help maintain the ability to provide credit under adverse conditions or liquidity requirements that mitigate susceptibility to abrupt capital outflows.

\textit{Second}, MPPs can help dampen procyclical dynamics triggered by capital inflows. The latter may result in unsustainable increases in credit, asset prices, unhedged foreign currency exposures, a further increase in cross-border non-core funding of the banking system and interconnectedness. The likelihood of surging gross capital inflows leading to a systemic crisis is considerably higher when funded by other investment (OI) flows \citep{hahm2013noncore}. 
The funding of substantial credit booms, mostly in foreign currency, via bank flows was particularly pronounced in most of the CESEE countries in the run-up to the global financial crisis whereby a large share of gross capital inflows involved flows from large global banks to their local subsidiaries. There is also a strong correlation between capital flows and the share of foreign currency lending to households, non-financial corporations and banks in almost all of the CESEE countries.

It follows that an MPP tightening shock may not only negatively affect credit extension to households and non-financial corporations, but might also have an effect on capital flows, while impacting different flow categories in different ways. OI inflows are expected to decline, insofar as MPP measures restrict bank lending in local and/or foreign currency. As a consequence, cross-border funding from parent banks, or generally from financial institutions and financial investors abroad, are less needed for domestic credit extension. Nevertheless, the case that OI inflows do not react negatively to an MPP tightening shock could be interpreted in such a way that the measures were not effective in reining in an overly excessive inflow of OI. But even if the impact of MPPs on bank inflows is negative, the effectiveness of the macroprudential tools can be limited by direct cross-border borrowing, as shown by \citet{ceruttizhou2018}. This was particularly the case prior to the GFC, when large foreign banks extended direct cross-border credit to non-financial corporates in some of the CESEE countries.  In general, lender-based MPPs are not effective to rein in excessive leverage in the economy, whereas borrower-based measures seem to be more appropriate, insofar that capital flows are not intermediated by banks. Concerning the impact of MPP measures on credit extension to households and non-financial corporations, it might also be the case that the effects are limited -- in particular during excessive capital inflows that allow banks to generate capital through retained earnings or issuing own capital \citep{basten2017higher}.

Apart from  total gross capital flows, the main focus in this study is on OI flows, as they typically have the most robust relationship with credit growth \citep{blanchard2016capital}. Restraining credit growth should affect bank inflows negatively, as argued above. The impact of macroprudential tightening on  other types of capital flows is, however, a priori ambiguous. FDIs related to the financial sector may decline as a consequence of stronger macroprudential regulation. Conversely, the need for capitalizing local subsidiaries of foreign banks may have a positive impact on FDI inflows. Concerning portfolio inflows, firms may substitute borrowing from banks as a consequence of MPP measures with issuing corporate bonds that are sold abroad \citep[recall the results of][mentioned in the introduction]{ahnert-etal-2018}.
To the extent that wholesale funding of banks is contained by MPP measures, the issuance of bank bonds (that are probably sold to foreign investors) may decline, yielding a negative impact on portfolio inflows. 

\section{Dynamic structural responses to macroprudential tightening} \label{sec:results}
Given the setup of the model and the variable definitions discussed in previous sections, we have a large number of possible combinations of results. Our findings may vary when examining different specifications of the FAVAR model (linear, nonlinear, different types of regime switching) or different types of capital flows. To get a robust picture across these various combinations, we summarize the main results for the responses of credit growth and capital inflows to the identified macroprudential tightening shock. We start by illustrating the impact of MPP tightening using a linear specification of the FAVAR, i.e. a specification without a regime switch (section~\ref{main:single}). The nonlinear results based on the endogenously specified regime shift are provided for both the high- and low-interest rate regime in section~\ref{main:ms} and appendix~\ref{add:MSirfs}. Additionally, appendix~\ref{add:detregime} provides results for the model with a deterministic regime allocation set equal to the onset of the GFC. Finally, section~\ref{main:summary} summarizes the results obtained across the different model specifications in view of our main research questions. Notwithstanding the nonlinear setup that allows for assessing whether responses to MPP shocks are different in particular subperiods, it should be noted that our impulse-response analysis is symmetric. Therefore, any finding for macroprudential policy tightening  holds -- with reverse sign -- also for macroprudential policy easing. By extension, the results shown in this section also illustrate the extent to which MPPs can mitigate the fallout of the COVID-19 pandemic. 

\subsection{Results from a linear FAVAR specification}\label{main:single}

\begin{figure}[!h]
	\centering
	\includegraphics[width=\linewidth]{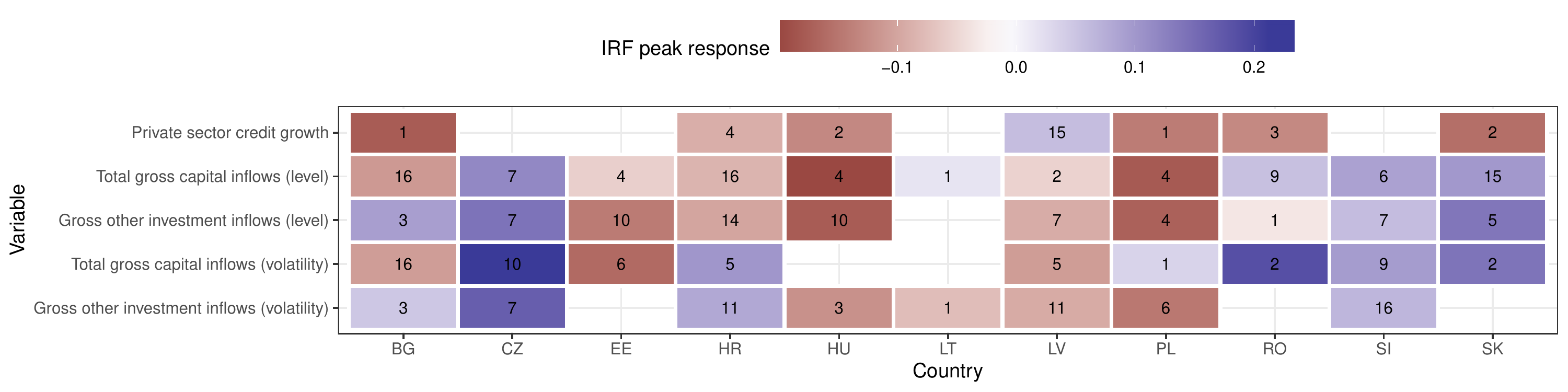}
\begin{minipage}{\textwidth}
\footnotesize \textbf{Note:} Red shaded cells denote negative and blue shaded cells denote positive responses. Cell numbers indicate the quarter after the shock at which the response reaches its peak. Empty cells refer to insignificance with respect to the $68$\,\% credible interval.  
\end{minipage}
	\caption{Entire-period (posterior median) \textbf{peak responses of private sector credit growth, total gross capital and gross other investment inflows} (levels \& volatilities) to a 1 SD tightening shock in the MPPI.}
	\label{fig:IRF_entire_peak_mapru}
\end{figure}

In this section, we start by considering a model setup that assumes the transmission mechanisms outlined in the previous section to be constant over time. To this end, we estimate a linear FAVAR model (i.e. $S_t=0, \forall t$) and study the impulse-response functions (IRFs) over the entire sample period. The peak IRFs to a 1 SD tightening shock in the baseline MPP indicator are summarized in Fig.~\ref{fig:IRF_entire_peak_mapru}. Empty cells indicate insignificant IRFs in the sense that the 68\,\% credible interval covers zero. The number in the box marks the quarter after the shock at which the posterior mean of the IRF reaches its minimum (negative numbers in red) or maximum (positive numbers in blue)\footnote{Note that the decision criterion for selecting the peak is always the absolute maximum value (i.e. in the few cases of changes from negative to positive responses -- or vice versa --, we take the larger one).}. 

Considering this, Fig.~\ref{fig:IRF_entire_peak_mapru} shows that private sector credit growth reacts negatively to tighter macroprudential regulations in six out of the eleven countries. These peak reactions appear to materialize within the first year after the shock hits the system, pointing towards a relatively fast transmission of MPPs to credit growth. These results resemble findings reported in \cite{meeks2017capital}, who provides VAR-based evidence for a decline in private and corporate lending after a MPP shock in the UK. Similarly, in a series of papers \cite{kim2017managing,kim2018effects,kim2019examining} and \cite{kim2019macroprudential} utilize a Panel VAR structure to obtain aggregate impulse responses for several Asian economies and show that tightening MPP shocks are indeed effective with regard to its primary goal of reducing private sector credit extension. Their findings also hold for several different subsets of countries who are differing in their employed exchange rate regime or other structural characteristics. 

Turning to the peak reactions of capital flows reveals that tighter MPPs result in decreasing levels of gross capital inflows in six out of the eleven countries under consideration. As opposed to the reactions of credit growth, these effects appear to be often longer-lived and tend to fade out much later. For instance, we observe that total gross capital inflows exhibit the strongest reactions four years after the shock in the case of Bulgaria and Croatia, while an earlier peak reaction, namely just after one year, can be observed in Estonia, Hungary and Poland. The responses of other investment inflows are mostly similar, except for Bulgaria whose peak response turns positive.
The volatility reactions, as shown in the last two rows of Fig.~\ref{fig:IRF_entire_peak_mapru}, are more diverse. For some countries, we find that MPPs succeed in lowering the volatility of capital inflows, whereas for other economies we observe that unexpected innovations to the MPP indicator increase capital flow volatility. However, it should be noted that in most countries where the capital flow levels showed already negative peak responses, their volatilities do so too (with the exception of Croatia).

\begin{figure}[!h]
	\centering
	\includegraphics[width=0.9\linewidth]{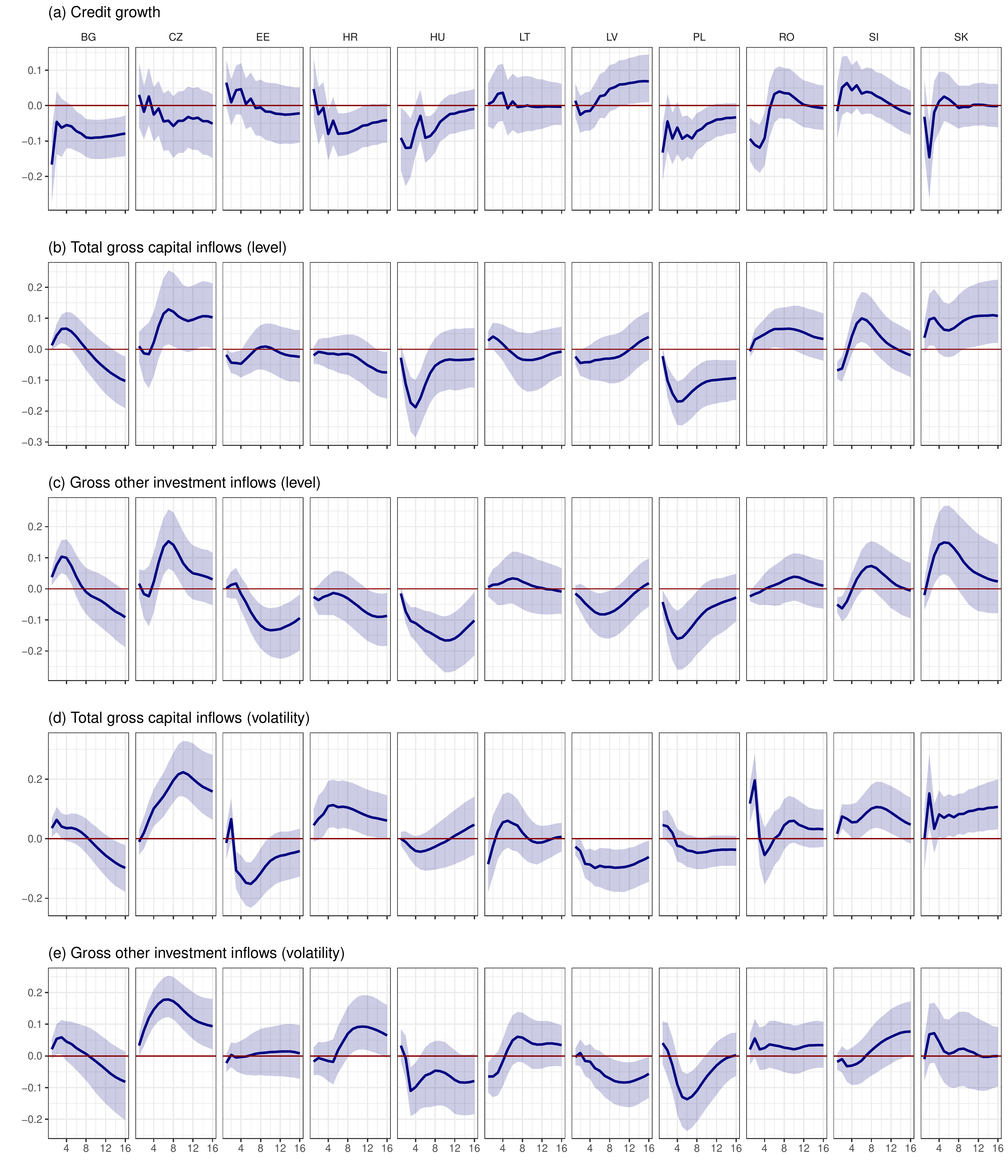}
\begin{minipage}{\textwidth}
\footnotesize \textbf{Note:} The blue line denotes the posterior median and blue shaded areas refer to the $68$\,\% credible set.  
\end{minipage}
	\caption{\textbf{Entire-period impulse-responses of private sector credit growth, total gross capital and gross other investment inflows} (levels \& volatilities) to a 1 SD tightening shock in the MPPI.}
	\label{fig:IRF_entire}
\end{figure}

For a closer inspection of the size, shape and evolution of the underlying IRFs, Fig.~\ref{fig:IRF_entire} provides their posterior median (blue line) surrounded by the corresponding 68\,\% credible interval (blue shaded area). Consistent with the discussion of the peak effects, we find that the impact of MPPs yields economically important effects. The reactions of private sector credit growth (shown in panel (a)) points towards strong cross-country heterogeneity in terms of shape and magnitudes. Some responses appear to be rather persistent (see, e.g., the reactions in Bulgaria and Poland) whereas other reactions are rather short-lived (see, e.g., Romania and Slovakia). In most cases, we observe that the impact of MPPs on credit growth is of transitory nature, fading out after several quarters.
Turning to the reaction of capital inflows in panels (b) and (c), we find that in a few cases there are economically very meaningful and persistent negative responses in Hungary and Poland. Sizable negative responses, though for only shorter periods, can be detected in the case of total capital inflows in Bulgaria and Croatia and in the case of other investment inflows in Estonia, Croatia and Latvia.
The volatility IRFs shown in panels (d) and (e) reveal quite some diversity across countries, with some economies responding to a MPP tightening with a pronounced decline in capital flow volatility (e.g. Latvia, Poland or Estonia, the latter regarding total flows), while a few others experience a sharp increase (e.g. the Czech Republic and Romania).



\clearpage

\subsection{Impulse-responses using a nonlinear modeling approach}\label{main:ms}

\begin{figure}[!htbp]
    \begin{minipage}{0.33\textwidth}
    \centering 
    (a) BG
    \end{minipage}
        \begin{minipage}{0.33\textwidth}
    \centering 
    (b) CZ
    \end{minipage}
    \begin{minipage}{0.33\textwidth}
    \centering 
    (c) EE
    \end{minipage}
	\begin{minipage}{0.33\textwidth}
	\centering
	\includegraphics[width=\linewidth]{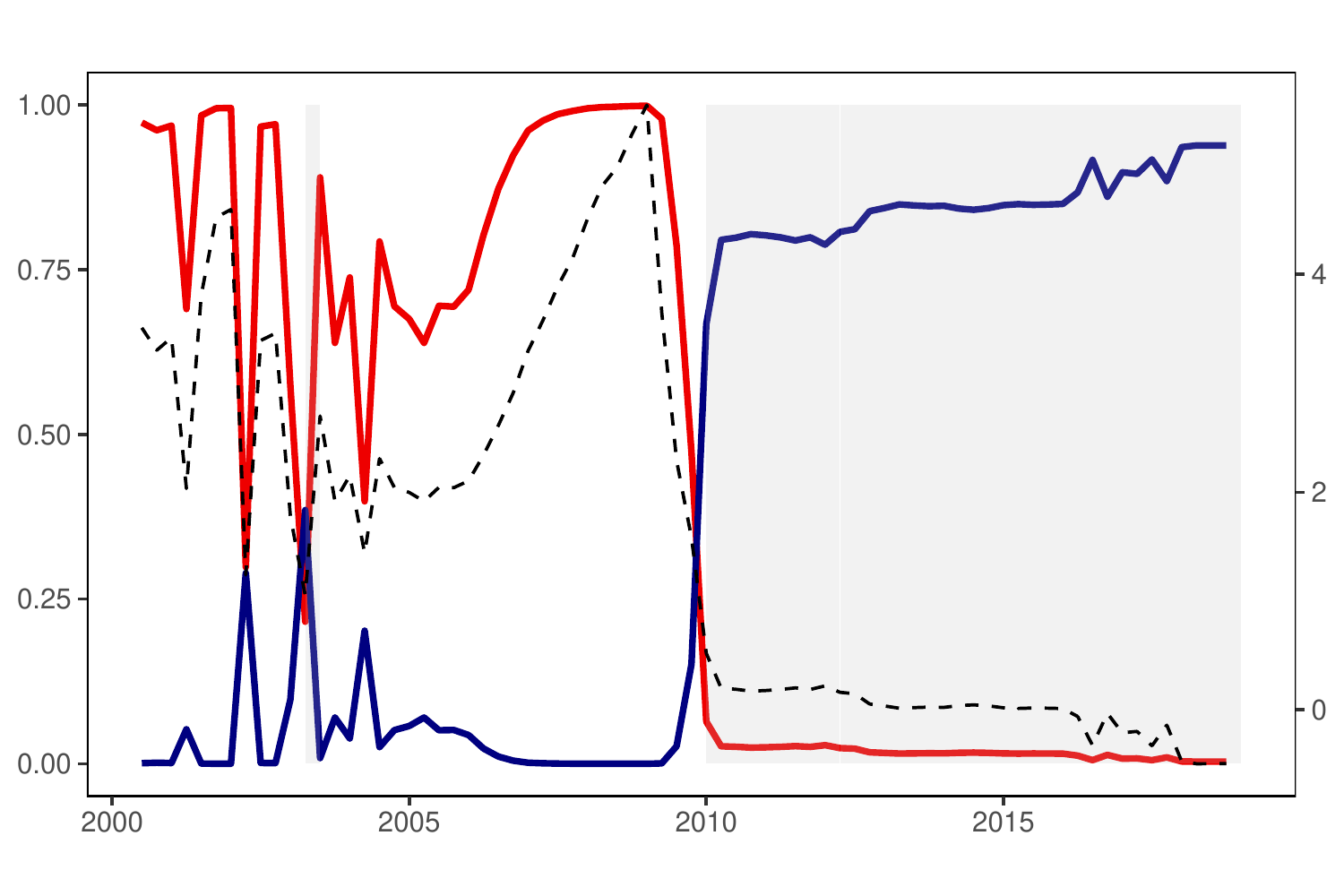}
	\end{minipage}
	\begin{minipage}{0.33\textwidth}
	\centering
	\includegraphics[width=\linewidth]{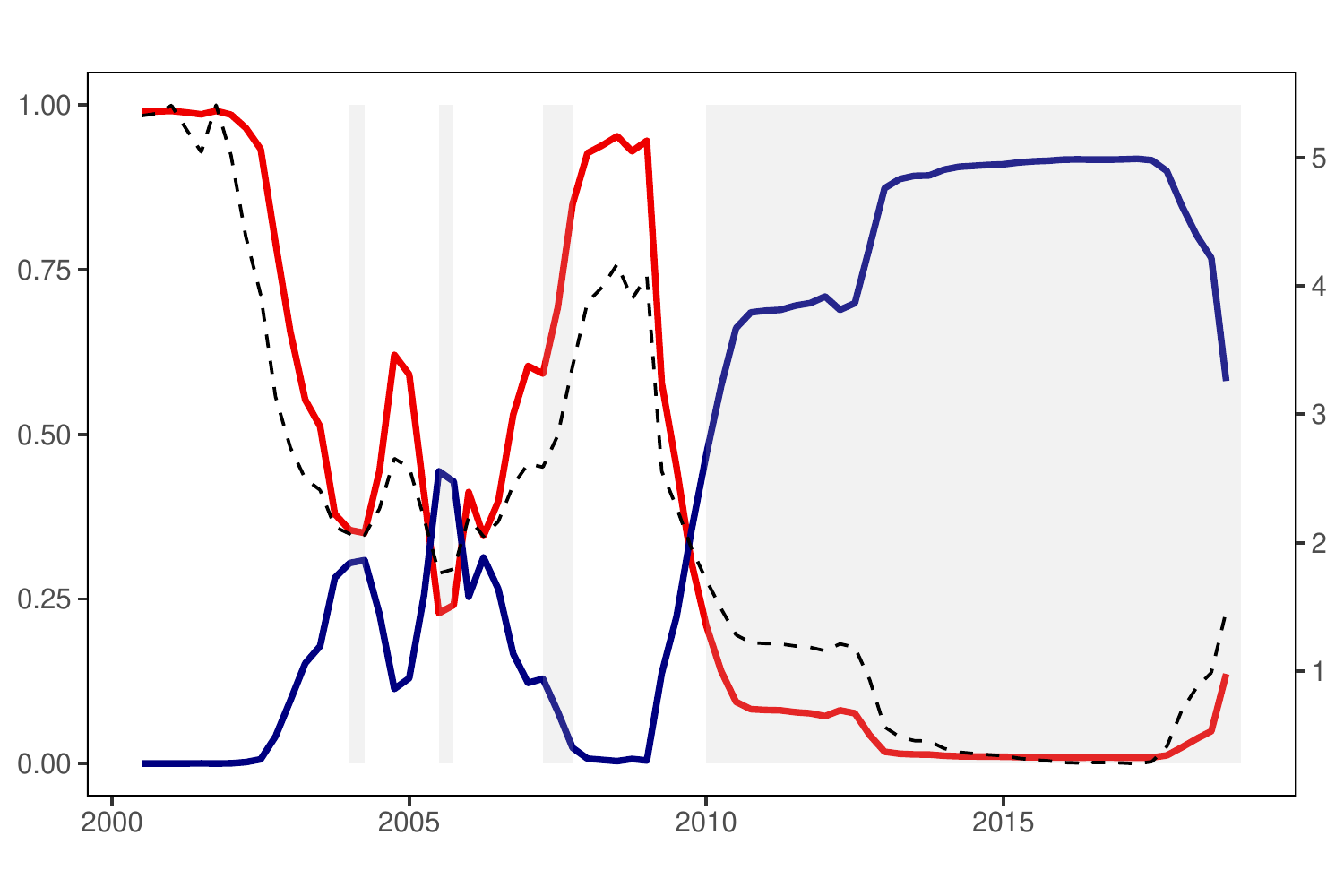}
	\end{minipage}
		\begin{minipage}{0.33\textwidth}
	\centering
	\includegraphics[width=\linewidth]{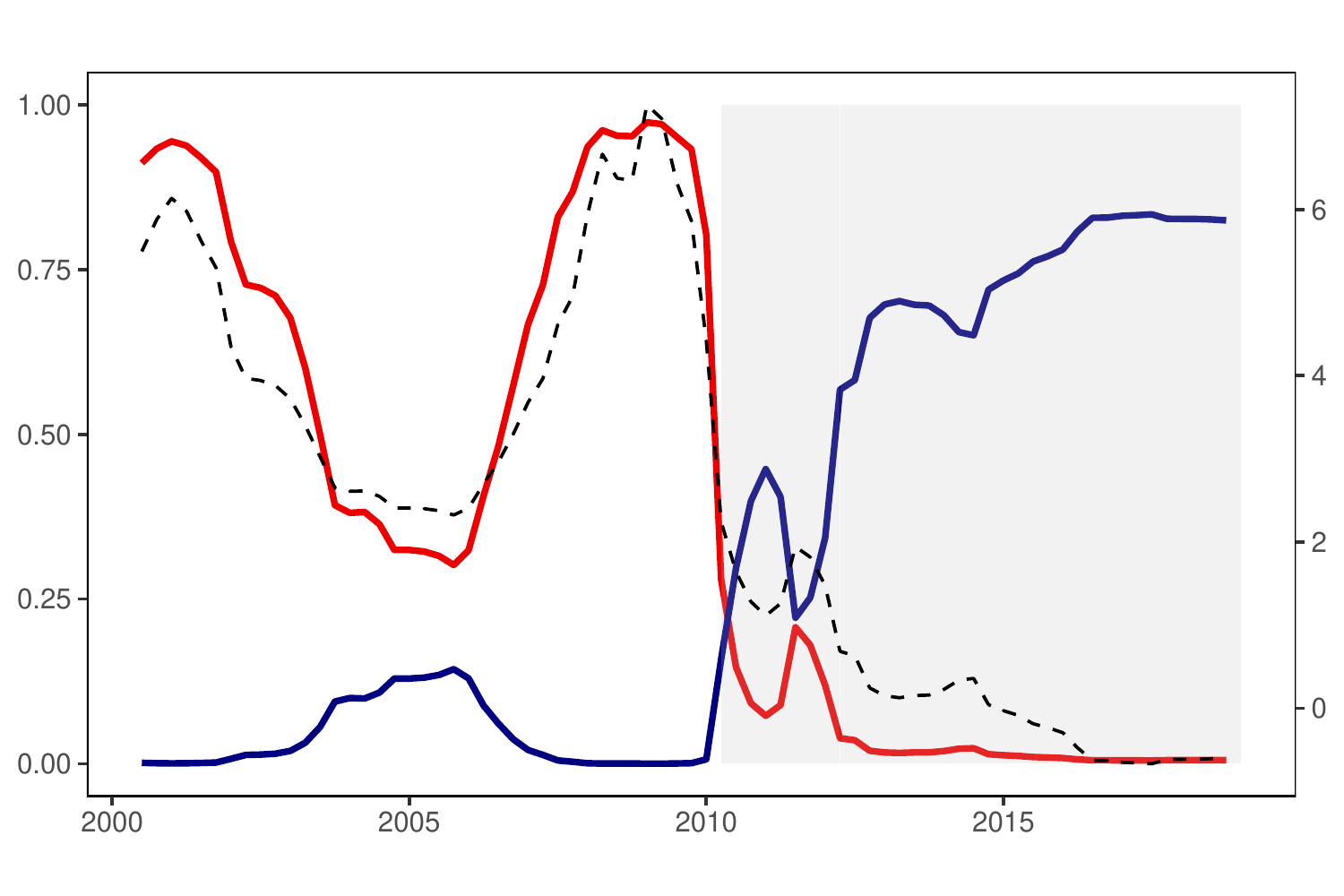}
	\end{minipage}	
	
    \begin{minipage}{0.33\textwidth}
    \centering 
    (d) HR
    \end{minipage}
        \begin{minipage}{0.33\textwidth}
    \centering 
    (e) HU
    \end{minipage}
    \begin{minipage}{0.33\textwidth}
    \centering 
    (f) LT
    \end{minipage}
	\begin{minipage}{0.33\textwidth}
	\centering
	\includegraphics[width=\linewidth]{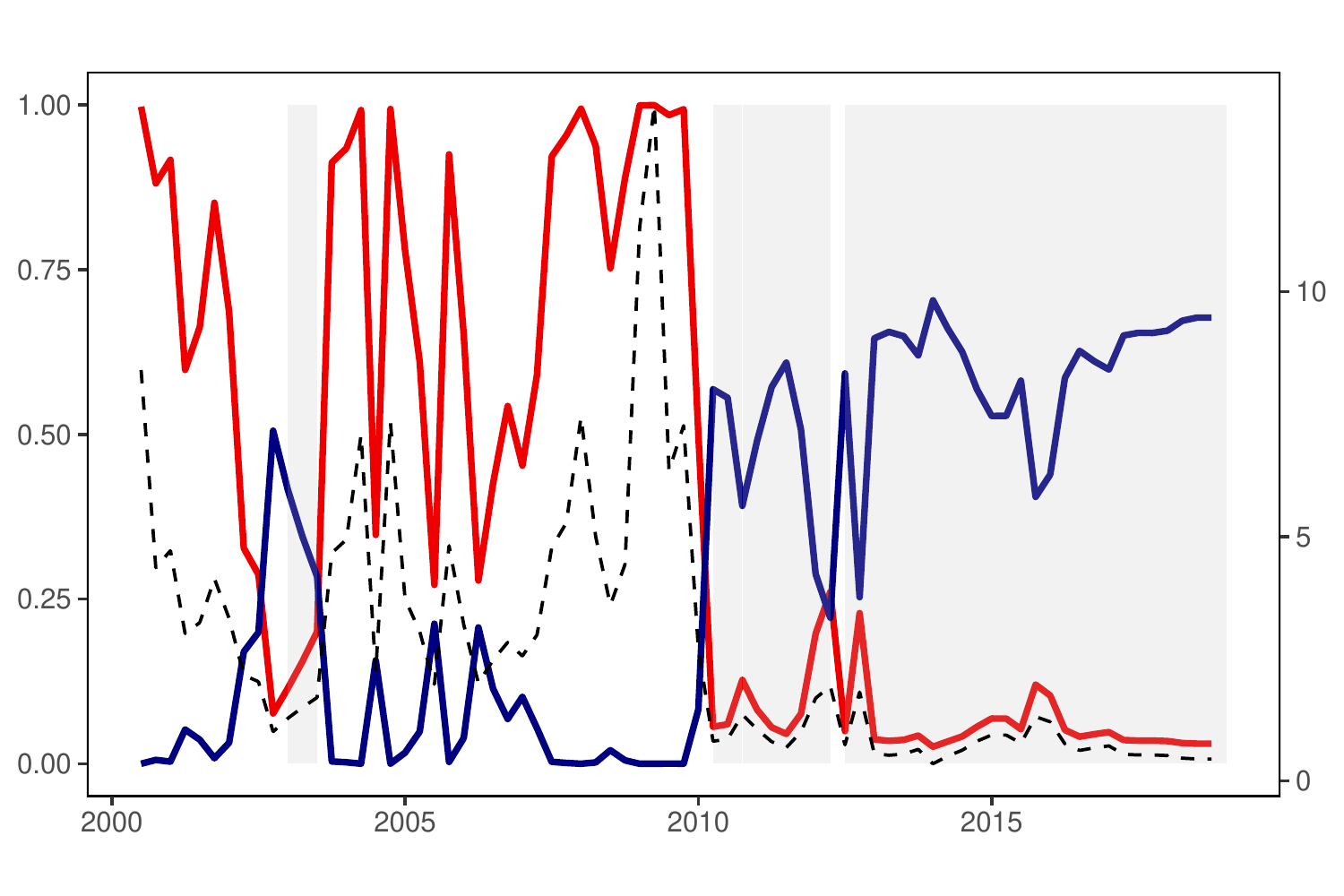}
	\end{minipage}
	\begin{minipage}{0.33\textwidth}
	\centering
	\includegraphics[width=\linewidth]{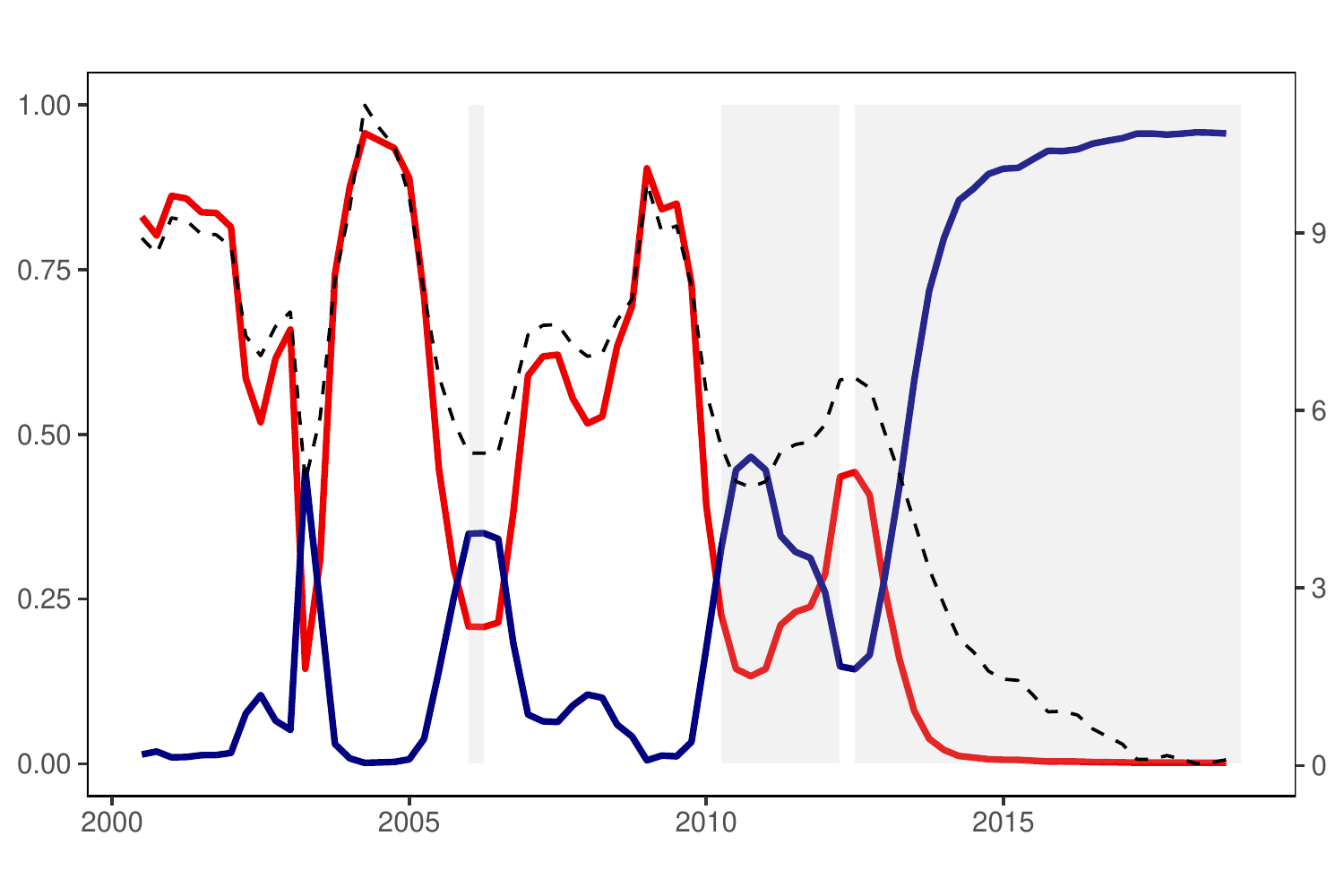}
	\end{minipage}
		\begin{minipage}{0.33\textwidth}
	\centering
	\includegraphics[width=\linewidth]{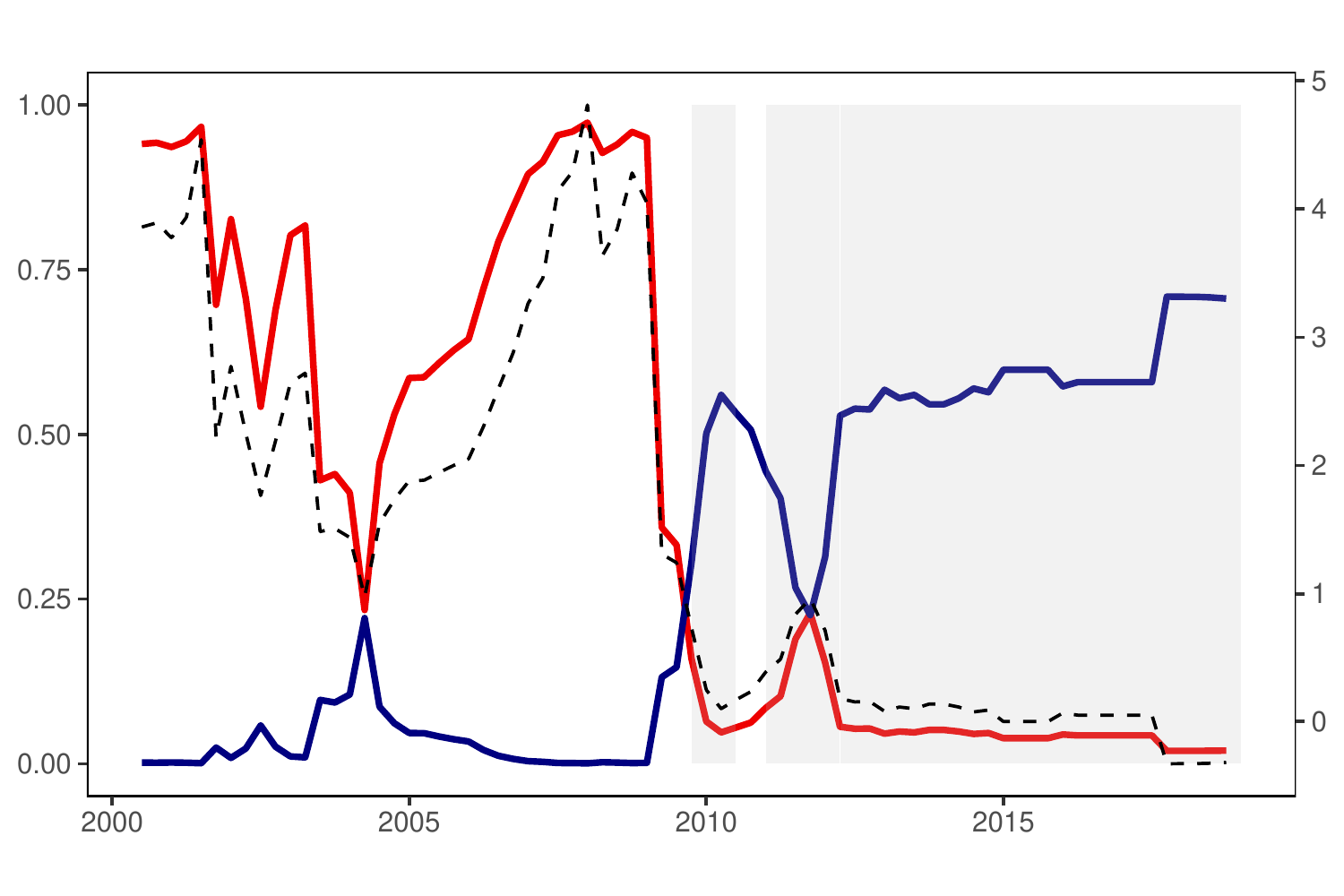}
	\end{minipage}		
	
	\begin{minipage}{0.33\textwidth}
    \centering 
    (g) LV
    \end{minipage}
        \begin{minipage}{0.33\textwidth}
    \centering 
    (h) PL
    \end{minipage}
    \begin{minipage}{0.33\textwidth}
    \centering 
    (i) RO
    \end{minipage}
	\begin{minipage}{0.33\textwidth}
	\centering
	\includegraphics[width=\linewidth]{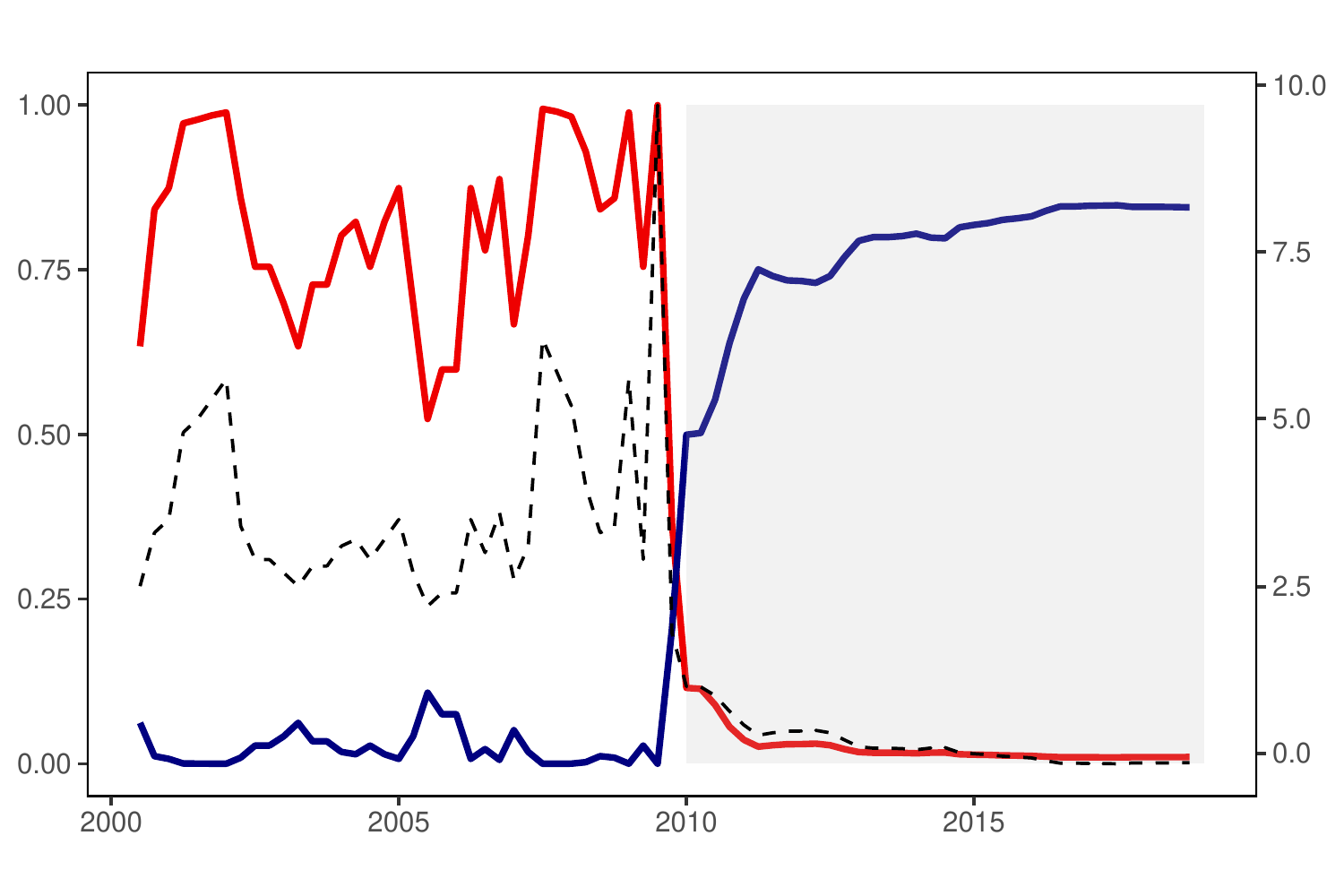}
	\end{minipage}
	\begin{minipage}{0.33\textwidth}
	\centering
	\includegraphics[width=\linewidth]{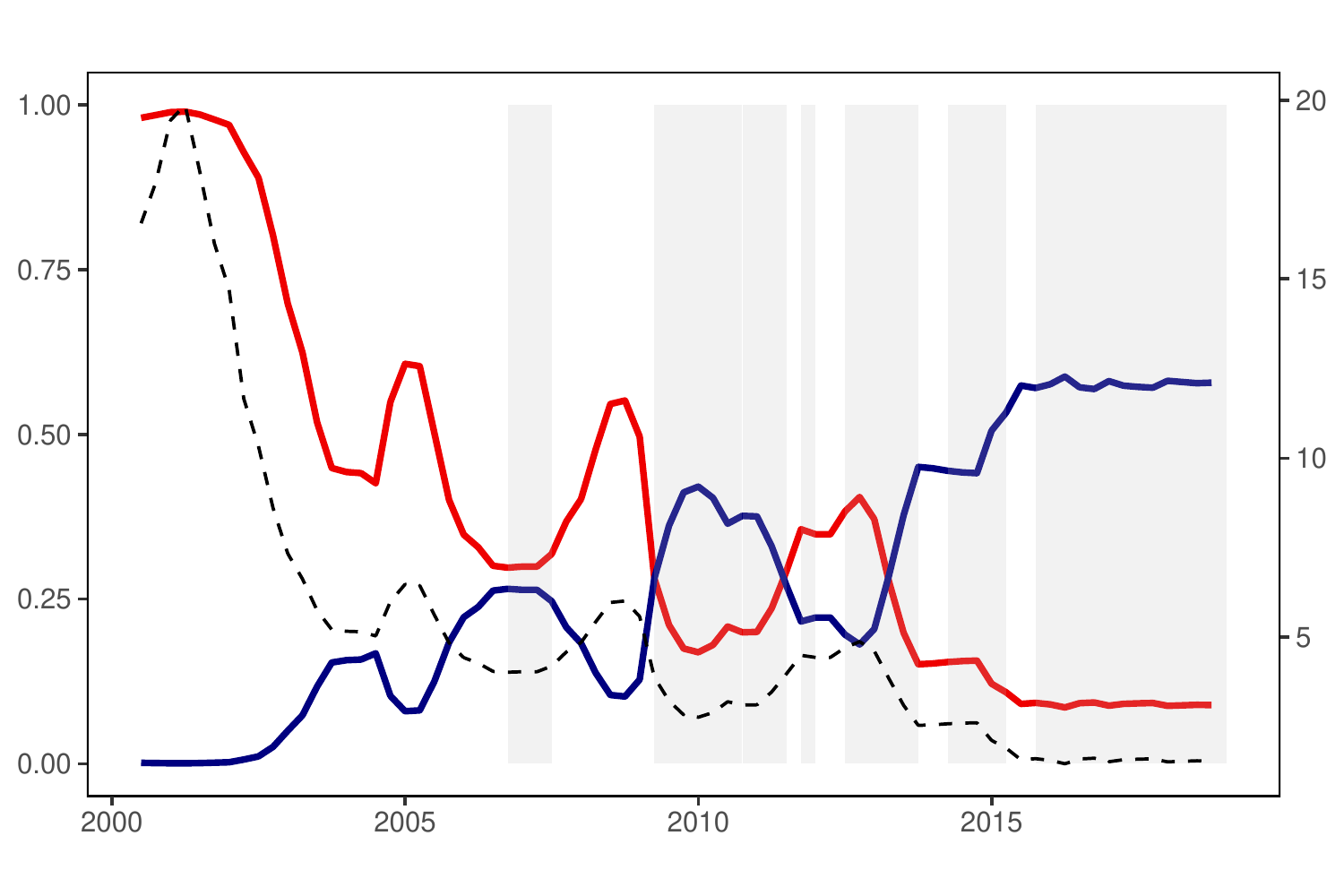}
	\end{minipage}
		\begin{minipage}{0.33\textwidth}
	\centering
	\includegraphics[width=\linewidth]{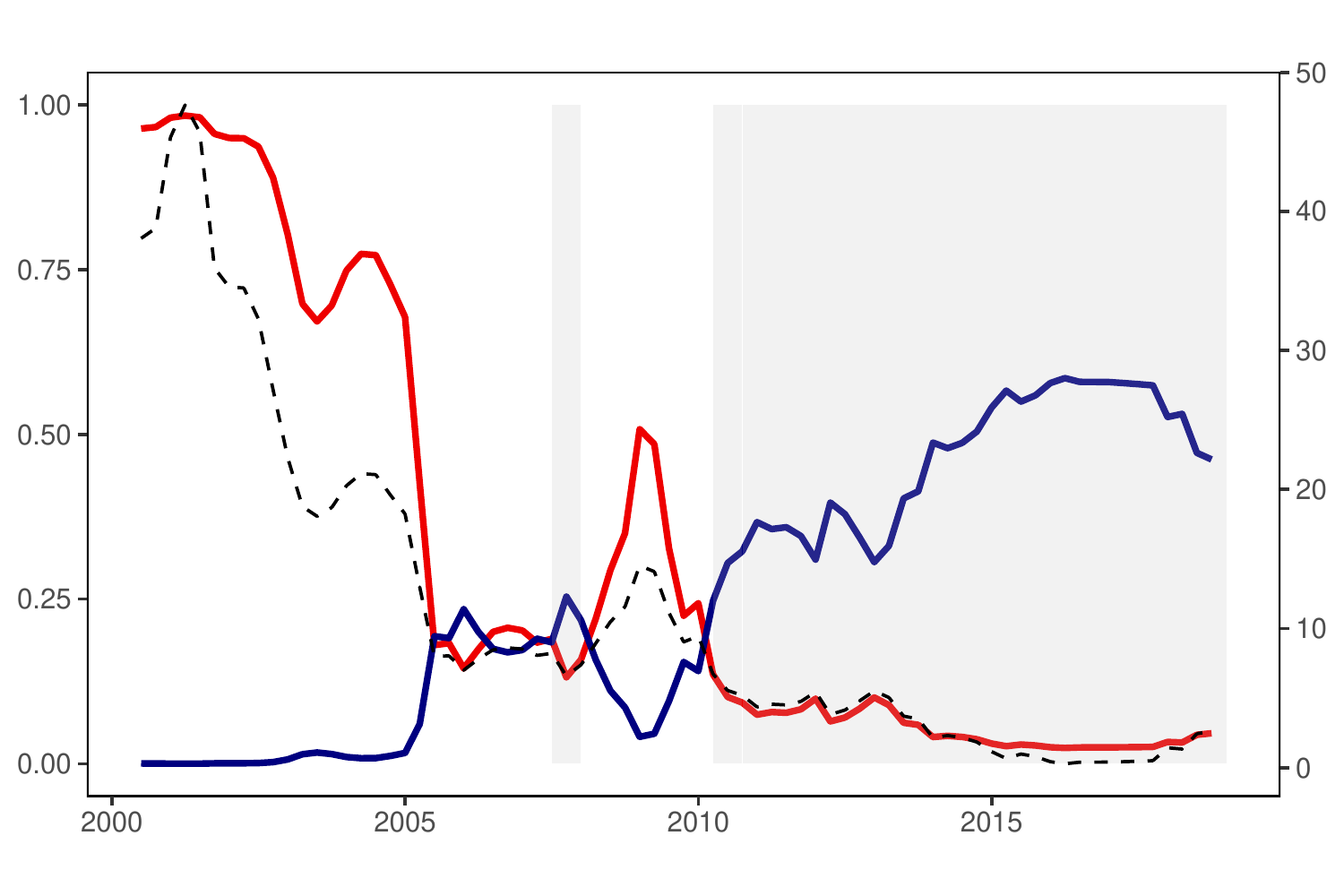}
	\end{minipage}		
	\begin{minipage}{0.33\textwidth}
    \centering 
    (j) SI
    \end{minipage}
    \begin{minipage}{0.33\textwidth}
    \centering 
    (k) SK
    \end{minipage}
    \begin{minipage}{0.33\textwidth}
    \centering 
    $ $
    \end{minipage}
	\begin{minipage}{0.33\textwidth}
	\centering
	\includegraphics[width=\linewidth]{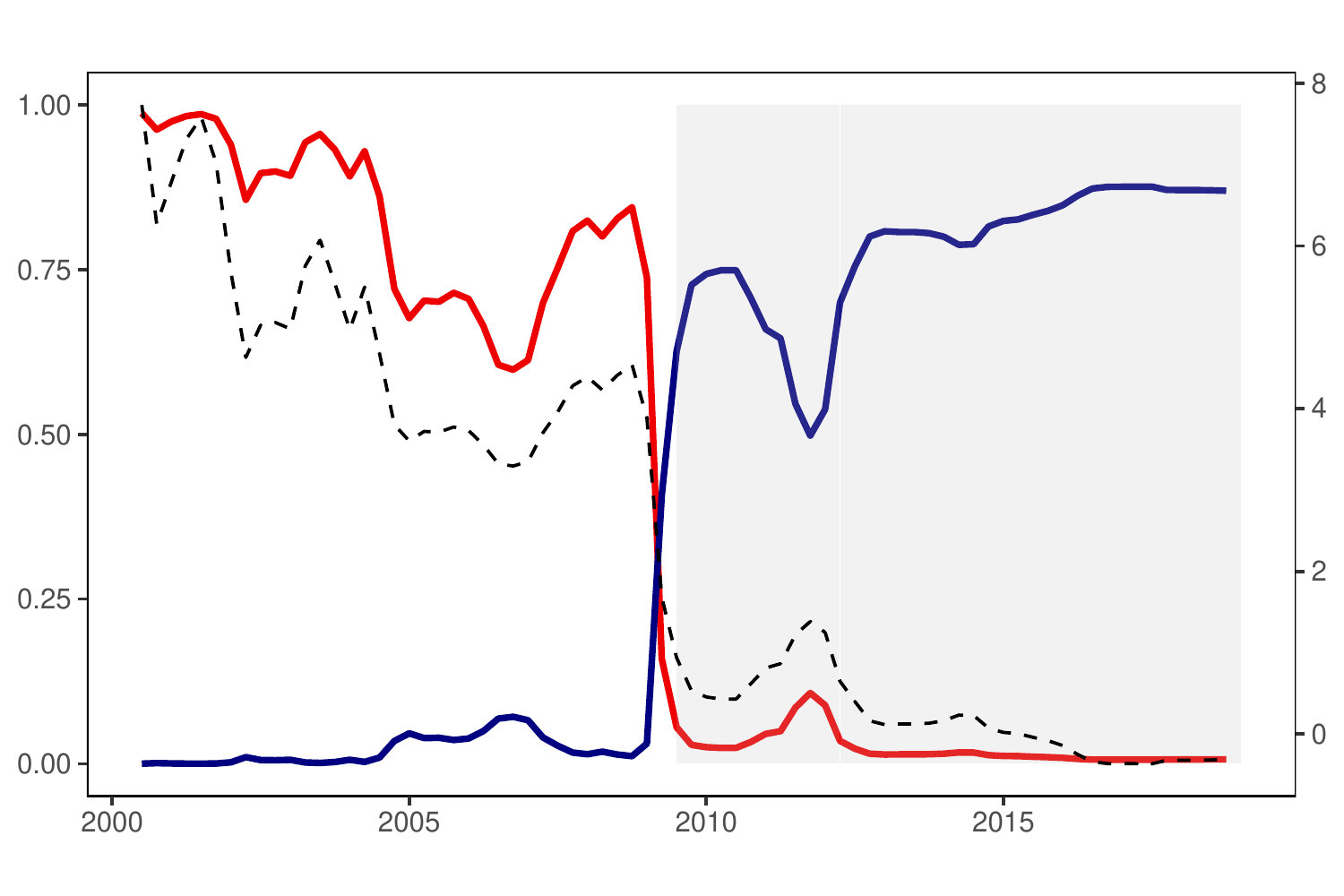}
	\end{minipage}
	\begin{minipage}{0.33\textwidth}	
	\centering
	\includegraphics[width=\linewidth]{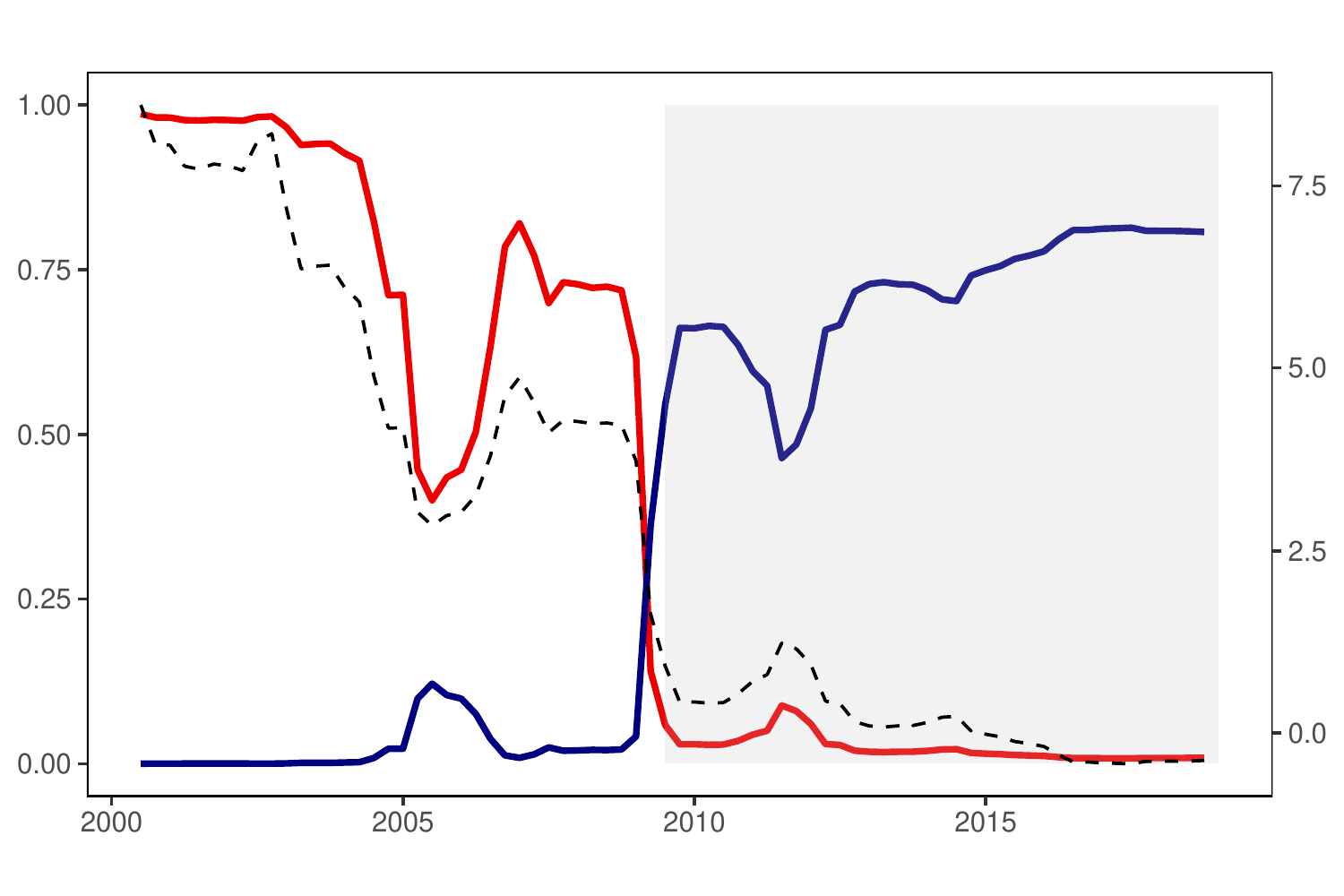}
	\end{minipage}	

\begin{minipage}{\textwidth}
\footnotesize \textbf{Note:} The red line indicates $\Pr(S_{it}=0|S_{it-1}=1)$, the blue line refers to $\Pr(S_{it}=1|S_{it-1}=0)$. The gray shaded areas indicate the posterior mean probability of a low-interest rate regime. The dashed black line depicts the country-specific interest rate. The left-hand axis shows the transition probabilities and the right-hand scale the values of interest rates.
\end{minipage}
	\caption{Posterior mean of transition probabilities and filtered probabilities of being in the high-interest rate regime $S_{it} = 0$ or in a low-interest rate regime $S_{it} = 1$.}
	\label{fig:TP_ms}
\end{figure}

In the previous section, we emphasized that MPP tightening leads to pronounced reactions of the quantities under consideration over the whole sample period. Next, we assess whether these reactions are different in specific subperiods, as the role of different transmission mechanisms may change over time. Before discussing the related IRFs, we focus on some features of our nonlinear modeling approach, such as the transition probabilities and the corresponding regime allocation, shown in Fig.~\ref{fig:TP_ms}.

Fig.~\ref{fig:TP_ms} shows by country the posterior mean of the transition probabilities (solid lines), the short-term interest rate (dashed) and the filtered probabilities (gray shaded areas) for being in a certain interest rate regime. From this figure, we can observe that most countries under investigation were quite uniformly considered to be part of the high-interest rate regime in the period before and shortly after the onset of the GFC (until about end-2009). This consistent pattern carries over to the post-GFC period. In accordance with the fact that most countries decreased their policy rates markedly to alleviate the impact of the GFC, a pronounced increase in the probability of moving into the low-interest rate regime can be observed. In parallel, we observe that the filtered probabilities of a given country being in the low-interest rate regime also tick up substantially. A notable exception is Poland, which shows more frequent fluctuations between the two regimes around the onset of the GFC and afterwards.

From this discussion, we observe that our model is quite successful in detecting regimes that are characterized by relatively low interest rates. Notice, however, that these regime allocations are stochastic, implying that our flexible specification is able to adapt the regime allocation accordingly, if other nonlinear features in our data set are observed.
Moreover, one additional takeaway from this discussion is that transition probabilities tend to feature time-variation, suggesting that lagged policy rates tend to predict movements in the regime allocation.

\begin{figure}[!htbp]
	\centering
	\includegraphics[width=\linewidth]{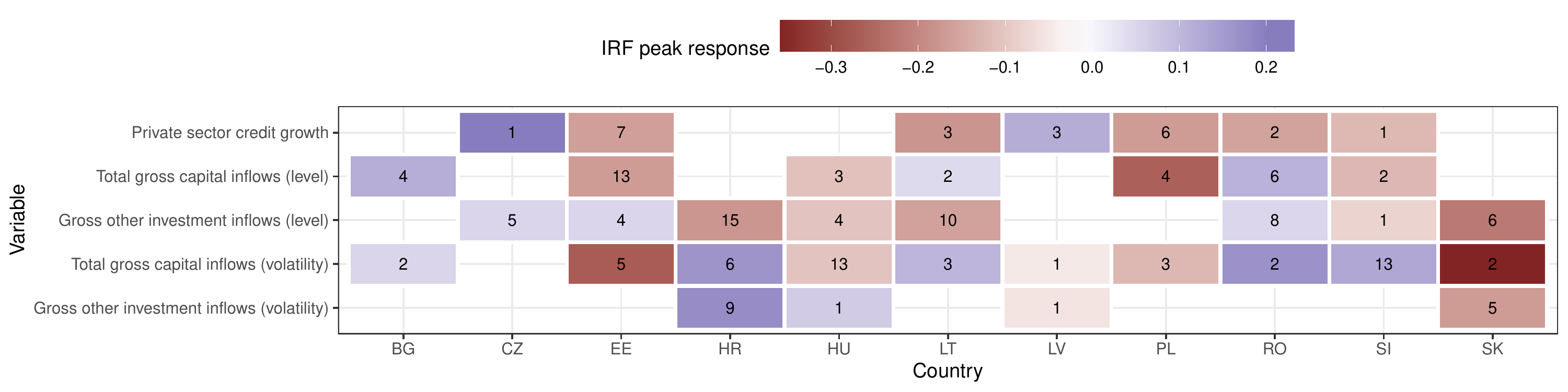}
\begin{minipage}{\textwidth}
\footnotesize \textbf{Note:} Red shaded cells denote negative and blue shaded cells denote positive responses. Cell numbers indicate the quarter after the shock at which the response reaches its peak.  Empty cells refer to insignificance with respect to the $68$\,\% credible interval.  
\end{minipage}
	\caption{High-interest rate regime (posterior median) \textbf{peak responses of private sector credit growth, total gross capital and gross other investment inflows} (levels \& volatilities) to a 1 SD tightening shock in the MPPI.}
	\label{fig:peak_high_ms}
\end{figure}

\begin{figure}[!htbp]
	\centering
	\includegraphics[width=\linewidth]{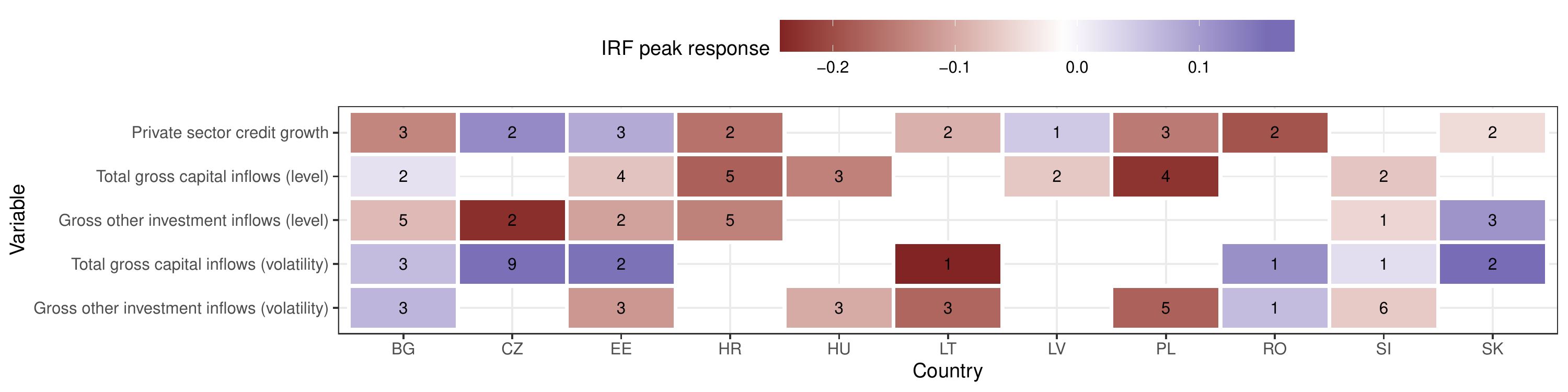}
\begin{minipage}{\textwidth}
\footnotesize \textbf{Note:} Red shaded cells denote negative and blue shaded cells denote positive responses. Cell numbers indicate the quarter after the shock at which the response reaches its peak. Empty cells refer to insignificance with respect to the 68\,\% credible interval.  
\end{minipage}
	\caption{Low-interest rate regime (posterior median) \textbf{peak responses of private sector credit growth, total gross capital and gross other investment inflows} (levels \& volatilities) to a 1 SD tightening shock in the MPPI.}
	\label{fig:peak_low_ms}
\end{figure}

In the next step, we again consider the dynamic reactions of our variables of interest to an MPP tightening shock. Figs.~\ref{fig:peak_high_ms} and~\ref{fig:peak_low_ms} report the peak responses of credit growth, capital inflow levels and volatilities for the high- and the low-interest rate regime, respectively. The corresponding Figs.~\ref{fig:IRF_high_ms_level} and~\ref{fig:IRF_low_ms_level} in appendix~\ref{add:MSirfs}, similar to Fig.~\ref{fig:IRF_entire}, again give an overview of the shape and evolution of the IRFs. 

In the high-interest rate regime, predominantly negative peak responses of credit growth are observable. The same holds true for level responses of both types of capital inflows. Hence, MPPs in an environment with high interest rates appear to be effective with regard to reining in excessive credit and capital inflows. However, a mixed pattern can again be observed for the corresponding volatilities of the capital flow series. For total capital inflows, a reduction in levels seems to be accompanied by a reduction in the corresponding volatility, while for other investment inflows the contrary is the case. Regarding the latter, generally rather few countries exhibit significant responses of their inflow volatility. 

The results for the low-interest rate regime, as depicted in Fig.~\ref{fig:peak_low_ms}, may be of relevance for the current COVID-19 crisis situation, as most countries are still, or again, subject to a low-interest rate environment. Under this regime, credit growth in Bulgaria, Croatia, Lithuania, Poland, Romania and Slovakia reacts negatively and these reactions occur in the first two or three quarters after the tightening shock in the macroprudential environment. Even more pronounced than in the high-interest rate regime, total capital inflows preponderantly show decreasing level reactions to an MPP tightening shock. Compared to the high-interest rate regime and entire-period results, this transmission is swifter. For capital flow volatilities, again a mixed picture emerges. While we can often observe declining volatilities of other investment inflows in response to an MPP tightening shock, volatilities of total capital inflows do in fact increase significantly in a majority of countries.

\clearpage

\subsection{Summary of results}\label{main:summary}

Given the wide range of possible model specifications and to provide a more compact overview of the different results, Table~\ref{tab:sum} summarizes the number of CESEE countries with statistically significant peak responses considering all specifications and variables of interest and grouped by negative (column ``$<0$'') and positive ones (column ``$>0$''). 
Based on this summary and in light of the transmission channels discussed in section~\ref{sec:transmission}, we can provide answers to three policy-relevant research questions (RQs):


\begin{enumerate}
\item[RQ 1.] \textit{Have MPP measures been effective in containing the provision of credit to households and firms?}
\item[] In line with expectations and large parts of the literature, not all but a majority of countries responds to a macroprudential tightening with a decline in private sector credit growth, in fact comparatively quickly -- within one year after the shock. The negative response is somewhat stronger pronounced (and occurs quicker) in the low-interest rate regime, as compared to other reference periods. This would be a good signal in the current low-interest rate environment of the COVID-19 crisis, as an MPP easing -- with reverse sign -- would give a positive impetus to lending activities and thus contribute to crisis mitigation. 

\item[RQ 2.] \textit{Have MPP measures curbed gross capital inflow volumes, especially bank flows?}
\item[] Not all but a majority of countries show a negative response of capital flow volumes to MPP tightening, indicating that MPPs can contribute to bringing back capital inflows to more sustainable levels in the case of an overshooting and would thus contribute to stabilization. Notably, negative volume responses are rather equally pronounced for total and other investment inflows and more often significant in a low-interest rate environment and the period after the global financial crisis. Thus, macroprudential tightening in the post-GFC episode could have reinforced cross-border deleveraging effects. On the other hand, MPP instruments in the pre-GFC period were probably not yet strong, developed or targeted enough to decisively contribute to a reduction in capital inflows. 

\item[RQ 3.] \textit{To which extent have MPP measures managed to shield countries from capital flow volatility?}
\item[] Contrary to the expectation of declining capital flow volatility in response to macroprudential tightening, our results indicate a somewhat mixed pattern. Positive volatility responses\footnote{Positive volatility responses to MPP tightening could also be due to the introduction of new sets of MPP instruments over time (e.g. borrower-based instruments or capital buffers in recent years). These new tools might have created stronger adjustment pressure for market participants and could thus also have led to more frequent investment or disinvestment decisions with a positive impact on capital flow volatilities.} often dominate in the case of total capital inflows, while positive and negative volatility responses are rather equally pronounced in the case of other investment inflows. Acknowledging notable cross-country heterogeneity, it should be emphasized that negative volatility responses are especially pronounced for total capital inflows in a high-interest rate environment (or the pre-GFC period) and for other investment inflows in a low-interest rate environment (or the post-GFC period), respectively. Overall, tighter MPPs seem to shield CESEE countries from capital flow volatility in a few instances (especially in the case of bank flows in the period since the GFC), but not generally. 
\end{enumerate}

\begin{table}[!htbp]
	\caption{Number of countries with significant peak responses to a tightening shock in macroprudential policies}
	\centering
	 \setlength\extrarowheight{0pt}
	 \begin{tabular}{llllccc}
		\textbf{Variable} & \textbf{FAVAR setup} & \textbf{Regime switch} & \textbf{Reference period} & \textbf{$<0$} & \textbf{$>0$} & \textbf{Total}\\
		\hline
        \multirow{5}{2.8cm}{Private sector credit growth} & Linear & None & Entire period & 6 & 1 & 7\\
                & Nonlinear & Endogenous & High-interest rate periods & 5 & 2 & 7\\
                & Nonlinear & Endogenous & Low-interest rate periods & 6 & 3 & 9\\
                & Nonlinear & Deterministic & Pre-GFC period & 5 & 1 & 6\\
                & Nonlinear & Deterministic & Post-GFC period & 3 & 5 & 8\\
                & & & Average of different models & 5 & 2 & 7 \\
		\hline
		\multirow{5}{2.8cm}{Total gross capital inflows (level)} & Linear & None & Entire period & 6 & 5 & 11\\
                & Nonlinear & Endogenous & High-interest rate periods & 4 & 3 & 7\\
                & Nonlinear & Endogenous & Low-interest rate periods & 6 & 1 & 7\\
                & Nonlinear & Deterministic & Pre-GFC period & 1 & 4 & 5\\
                & Nonlinear & Deterministic & Post-GFC period & 5 & 2 & 7\\
                & & & Average of different models & 4 & 3 & 7 \\
		\hline
        \multirow{5}{2.8cm}{Gross other investment inflows (level)}  & Linear & None & Entire period & 6 & 4 & 10\\
                & Nonlinear & Endogenous & High-interest rate periods & 5 & 3 & 8\\
                & Nonlinear & Endogenous & Low-interest rate periods & 5 & 1 & 6\\
                & Nonlinear & Deterministic & Pre-GFC period & 3 & 2 & 5\\
                & Nonlinear & Deterministic & Post-GFC period & 6 & 2 & 8\\
                & & & Average of different models & 5 & 2 & 7 \\
		\hline
        \multirow{5}{2.8cm}{Total gross capital inflows (volatility)}  & Linear & None & Entire period & 3 & 6 & 9\\
                & Nonlinear & Endogenous & High-interest rate periods & 5 & 5 & 10\\
                & Nonlinear & Endogenous & Low-interest rate periods & 1 & 6 & 7\\
                & Nonlinear & Deterministic & Pre-GFC period & 4 & 3 & 7\\
                & Nonlinear & Deterministic & Post-GFC period & 4 & 5 & 9\\
                & & & Average of different models & 3 & 5 & 8 \\
		\hline
        \multirow{5}{2.8cm}{Gross other investment inflows (volatility)}  & Linear & None & Entire period & 4 & 4 & 8\\
                & Nonlinear & Endogenous & High-interest rate periods & 2 & 2 & 4\\
                & Nonlinear & Endogenous & Low-interest rate periods & 5 & 2 & 7\\
                & Nonlinear & Deterministic & Pre-GFC period & 1 & 2 & 3\\
                & Nonlinear & Deterministic & Post-GFC period & 5 & 3 & 8\\
                & & & Average of different models & 3 & 3 & 6 \\
		\hline
	\end{tabular}
	
	\vspace{1ex}
	
	{\raggedright \footnotesize \textbf{Note:} This table shows the number of significant negative vs. positive peak responses of selected variables to the identified tightening (1 SD) shock in the MPPI, based on nonlinear or linear FAVAR estimates, respectively, over the period 2000--2018 across the 11 CESEE EU Member States. Significance inference is based on 68\,\% credible sets. \par}
	\label{tab:sum}
\end{table}

\clearpage

\section{Conclusions}\label{sec:con}

Studying the impact of MPPs on capital flows in the CESEE countries is appealing for at least two reasons. \textit{First}, CESEE countries have experienced a substantial boom-bust cycle in capital flows. The corresponding credit cycle was pronounced, too. Because of the large reversal of flows (in particular related to bank flows) during the 2008/2009 crisis, the CESEE region suffered stronger output declines than any other region in the world  \citep{berglof2009understanding}. \textit{Second}, in contrast to the experience of advanced economies with MPPs, which attracted more attention only in the aftermath of the global financial crisis, some CESEE countries, e.g. Bulgaria, Croatia and Romania, had  been quite active in adopting MPPs already before the crisis -- on the back of extraordinary credit growth, predominantly denominated in foreign currency, at the time. 

Examining the impact of MPPs on domestic macroeconomic variables, such as the credit cycle, has gained a lot of attention in recent years. Still, there are only a few studies on the related global dimension \citep[as summarized in][]{ESRB-2020}, such as the role of international spillovers, cross-border leakages or, more generally, capital flow responses. The question whether MPPs are effective in taming capital flows is particularly interesting for the CESEE countries, as MPPs are expected to have had a sizable impact on cross-border flows -- in particular on bank flows -- and their volatility, given the prominent role of foreign-owned banks in the region.

To measure MPPs, we rely on an intensity-adjusted index by  \citep{eller-etal-taxonomy}, which allows for capturing both if and to what extent the respective MPP tool was implemented. Moreover, to study the dynamic responses of capital flows to MPP shocks, a novel regime-switching factor-augmented vector autoregressive (FAVAR) model has been applied. It allows for capturing potential structural breaks in the policy regime and controls -- besides domestic macroeconomic quantities -- for the impact of global factors such as the global financial cycle over the period from 2000 to 2018. The question of how MPPs contribute to shielding countries from fluctuations in the global financial cycle is answered by including an estimated measure of capital flow volatility in the dynamic econometric specification.

This empirical analysis reveals that tighter MPPs could apparently be effective in containing private sector credit growth and the volumes of gross capital inflows in a majority of the CESEE countries analyzed. Negative 
responses of credit growth and capital flow volumes are somewhat stronger pronounced (and occur quicker) in a low-interest rate environment or, in the case of capital flows, in the period after the global financial crisis. Finally, the responses of capital flow volatilities to an MPP tightening shock display a rather mixed pattern with both positive and negative responses being important. Tighter MPPs seem to shield CESEE countries from capital flow volatility in a few instances (especially in the case of bank flows in the period since the GFC), but not generally. 

It should be noted that a few countries deviate from these general patterns. We conjecture that the reasons  for cross-country heterogeneity can be attributed to differences in the composition of MPPs, domestic financial cycles or the respective exchange rate regime -- issues that have to be explored further in subsequent research.

\clearpage

\clearpage
\small{\scfont\setstretch{0.85}
\addcontentsline{toc}{section}{References}
\bibliographystyle{fischer}
\bibliography{References}} 

\begin{thebibliography}{63}
\providecommand{\natexlab}[1]{#1}

\bibitem[{Ahnert et~al.(2018)Ahnert, Forbes, Friedrich and
  Reinhardt}]{ahnert-etal-2018}
Ahnert T, Forbes K, Friedrich C and Reinhardt D (2018) {Macroprudential FX
  regulations: shifting the snowbanks of FX vulnerability?}
\newblock NBER Working Paper 25083, The National Bureau of Economic Research

\bibitem[{Aizenman et~al.(2017)Aizenman, Chinn and Ito}]{aizenman2017financial}
Aizenman J, Chinn MD and Ito H (2017) Financial spillovers and macroprudential
  policies.
\newblock Working Paper 24105, National Bureau of Economic Research

\bibitem[{Akram(2014)}]{akram2014macro}
Akram QF (2014) Macro effects of capital requirements and macroprudential
  policy.
\newblock \emph{Economic Modelling} 42, 77--93

\bibitem[{Alam et~al.(2019)Alam, Alter, Eiseman, Gelos, Kang, Narita, Nier and
  Wang}]{alam2019digging}
Alam Z, Alter A, Eiseman J, Gelos RG, Kang H, Narita M, Nier E and Wang N
  (2019) Digging deeper -- evidence on the effects of macroprudential policies
  from a new database.
\newblock IMF Working Paper 19/66, International Monetary Fund

\bibitem[{Allenby et~al.(1998)Allenby, Arora and
  Ginter}]{allenby1998heterogeneity}
Allenby GM, Arora N and Ginter JL (1998) On the heterogeneity of demand.
\newblock \emph{Journal of Marketing Research} 35(3), 384--389

\bibitem[{Amisano and Fagan(2013)}]{amisano2013money}
Amisano G and Fagan G (2013) Money growth and inflation: A regime switching
  approach.
\newblock \emph{Journal of International Money and Finance} 33, 118--145

\bibitem[{Angelini et~al.(2014)Angelini, Neri and
  Panetta}]{angelini2014interaction}
Angelini P, Neri S and Panetta F (2014) The interaction between capital
  requirements and monetary policy.
\newblock \emph{Journal of Money, Credit and Banking} 46(6), 1073--1112

\bibitem[{Avdjiev et~al.(2018)Avdjiev, Bruno, Koch and
  Shin}]{avdjiev2018dollar}
Avdjiev S, Bruno V, Koch C and Shin HS (2018) The dollar exchange rate as a
  global risk factor: evidence from investment.
\newblock BIS Working Papers 695, Bank for International Settlements

\bibitem[{Aysan et~al.(2015)Aysan, Fendo{\u{g}}lu and Kilinc}]{aysan2014mpp}
Aysan AF, Fendo{\u{g}}lu S and Kilinc M (2015) Macroprudential policies as
  buffer against volatile cross-border capital flows.
\newblock \emph{The Singapore Economic Review} 60(01), 1550001

\bibitem[{Basten and Koch(2015)}]{basten2017higher}
Basten CC and Koch C (2015) Higher bank capital requirements and mortgage
  pricing: evidence from the Countercyclical Capital Buffer (CCB).
\newblock BIS Working Paper 511, Bank for International Settlements

\bibitem[{Beirne and Friedrich(2014)}]{beirne2014}
Beirne J and Friedrich C (2014) Capital flows and macroprudential policies -- a
  multilateral assessment of effectiveness and externalities.
\newblock ECB Working Paper 1721, European Central Bank

\bibitem[{Beirne and Friedrich(2017)}]{beirne2017}
Beirne J and Friedrich C (2017) Macroprudential policies, capital flows, and
  the structure of the banking sector.
\newblock \emph{Journal of International Money and Finance} 75, 47--68

\bibitem[{Bergant et~al.(2020)Bergant, Grigoli, Hansen and
  Sandri}]{bergant-etal-2020}
Bergant K, Grigoli F, Hansen NJ and Sandri D (2020) Dampening global financial
  shocks: can macroprudential regulation help (more than capital controls)?
\newblock IMF Working Paper 20/106, International Monetary Fund

\bibitem[{Bergl{\"o}f et~al.(2010)Bergl{\"o}f, Korniyenko, Plekhanov and
  Zettelmeyer}]{berglof2009understanding}
Bergl{\"o}f E, Korniyenko Y, Plekhanov A and Zettelmeyer J (2010)
  {Understanding the crisis in Emerging Europe}.
\newblock \emph{Public Policy Review} 6(6), 985--1008

\bibitem[{Bernanke et~al.(2005)Bernanke, Boivin and
  Eliasz}]{bernanke2005measuring}
Bernanke B, Boivin J and Eliasz P (2005) Measuring the effects of monetary
  policy: a factor-augmented vector autoregressive (FAVAR) approach.
\newblock \emph{The Quarterly Journal of Economics} 120(1), 387--422

\bibitem[{Billio et~al.(2016)Billio, Casarin, Ravazzolo and
  Van~Dijk}]{billio2016interconnections}
Billio M, Casarin R, Ravazzolo F and Van~Dijk HK (2016) Interconnections
  between eurozone and US booms and busts using a Bayesian panel
  Markov-switching VAR model.
\newblock \emph{Journal of Applied Econometrics} 31(7), 1352--1370

\bibitem[{Blanchard et~al.(2016)Blanchard, Ostry, Ghosh and
  Chamon}]{blanchard2016capital}
Blanchard O, Ostry JD, Ghosh AR and Chamon M (2016) Capital flows: expansionary
  or contractionary?
\newblock \emph{American Economic Review} 106(5), 565--69

\bibitem[{Calvo et~al.(1996)Calvo, Leiderman and
  Reinhart}]{calvoleidermanreinhart1996}
Calvo GA, Leiderman L and Reinhart CM (1996) Inflows of capital to developing
  countries in the 1990s.
\newblock \emph{Journal of Economic Perspectives} 10(2), 123--139

\bibitem[{Cerutti et~al.(2017)Cerutti, Claessens and Laeven}]{cerutti2017use}
Cerutti E, Claessens S and Laeven L (2017) The use and effectiveness of
  macroprudential policies: new evidence.
\newblock \emph{Journal of Financial Stability} 28, 203--224

\bibitem[{Cerutti and Zhou(2018)}]{ceruttizhou2018}
Cerutti E and Zhou H (2018) Cross-border banking and the circumvention of
  macroprudential and capital control measures.
\newblock IMF Working Paper 18/217, International Monetary Fund

\bibitem[{Cesa-Bianchi et~al.(2018)Cesa-Bianchi, Ferrero and
  Rebucci}]{cesa2018international}
Cesa-Bianchi A, Ferrero A and Rebucci A (2018) International credit supply
  shocks.
\newblock \emph{Journal of International Economics} 112, 219--237

\bibitem[{Coman and Lloyd(2019)}]{coman2019face}
Coman A and Lloyd SP (2019) In the face of spillovers: prudential policies in
  emerging economies.
\newblock ECB Working Paper 2339, European Central Bank

\bibitem[{Dumi{\v{c}}i{\'c}(2018)}]{dumivcic2018effectiveness}
Dumi{\v{c}}i{\'c} M (2018) Effectiveness of macroprudential policies in Central
  and Eastern European countries.
\newblock \emph{Public Sector Economics} 42(1), 1--19

\bibitem[{Eller et~al.(2016)Eller, Huber and
  Schuberth}]{eller2016understanding}
Eller M, Huber F and Schuberth H (2016) Understanding the drivers of capital
  flows into the CESEE countries.
\newblock \emph{Focus on European Economic Integration} Q2/16, 79--104

\bibitem[{Eller et~al.(2020{\natexlab{a}})Eller, Huber and
  Schuberth}]{eller-huber-schuberth-2020}
Eller M, Huber F and Schuberth H (2020{\natexlab{a}}) How important are global
  factors for understanding the dynamics of international capital flows?
\newblock \emph{Journal of International Money and Finance} forthcoming

\bibitem[{Eller et~al.(2020{\natexlab{b}})Eller, Martin, Schuberth and
  Vashold}]{eller-etal-taxonomy}
Eller M, Martin R, Schuberth H and Vashold L (2020{\natexlab{b}})
  Macroprudential policies in {CESEE} -- an intensity-adjusted approach.
\newblock \emph{Focus on European Economic Integration} Q2/20, 65--81

\bibitem[{Fendo{\u{g}}lu(2017)}]{fendouglu2017credit}
Fendo{\u{g}}lu S (2017) Credit cycles and capital flows: effectiveness of the
  macroprudential policy framework in emerging market economies.
\newblock \emph{Journal of Banking \& Finance} 79, 110--128

\bibitem[{Filardo(1994)}]{filardo1994business}
Filardo AJ (1994) Business-cycle phases and their transitional dynamics.
\newblock \emph{Journal of Business \& Economic Statistics} 12(3), 299--308

\bibitem[{Forbes et~al.(2015)Forbes, Fratzscher and Straub}]{forbes2015capital}
Forbes K, Fratzscher M and Straub R (2015) Capital-flow management measures:
  what are they good for?
\newblock \emph{Journal of International Economics} 96, S76--S97

\bibitem[{Fratzscher(2012)}]{fratzscher2012}
Fratzscher M (2012) Capital flows, push versus pull factors and the global
  financial crisis.
\newblock \emph{Journal of International Economics} 88(2), 341--356

\bibitem[{Frost et~al.(2020)Frost, Ito and van Stralen}]{frost-etal-2020}
Frost J, Ito H and van Stralen R (2020) The effectiveness of macroprudential
  policies and capital controls against volatile capital inflows.
\newblock BIS Working Papers 867, Bank for International Settlements

\bibitem[{Fr{\"u}hwirth-Schnatter et~al.(2004)Fr{\"u}hwirth-Schnatter,
  T{\"u}chler and Otter}]{fruhwirth2004bayesian}
Fr{\"u}hwirth-Schnatter S, T{\"u}chler R and Otter T (2004) Bayesian analysis
  of the heterogeneity model.
\newblock \emph{Journal of Business \& Economic Statistics} 22(1), 2--15

\bibitem[{Galati and Moessner(2013)}]{galati2013macroprudential}
Galati G and Moessner R (2013) Macroprudential policy -- a literature review.
\newblock \emph{Journal of Economic Surveys} 27(5), 846--878

\bibitem[{Gerali et~al.(2010)Gerali, Neri, Sessa and
  Signoretti}]{gerali2010credit}
Gerali A, Neri S, Sessa L and Signoretti FM (2010) Credit and banking in a DSGE
  model of the Euro area.
\newblock \emph{Journal of Money, Credit and Banking} 42, 107--141

\bibitem[{Hahm et~al.(2013)Hahm, Shin and Shin}]{hahm2013noncore}
Hahm Jh, Shin HS and Shin K (2013) Noncore bank liabilities and financial
  vulnerability.
\newblock \emph{Journal of Money, Credit and Banking} 45(S1), 3--36

\bibitem[{Hauzenberger and Huber(2020)}]{hauzenberger2020fx}
Hauzenberger N and Huber F (2020) Model instability in predictive exchange rate
  regressions.
\newblock \emph{Journal of Forecasting} 39(2), 168--186

\bibitem[{Huber and Fischer(2018)}]{huber2018markov}
Huber F and Fischer MM (2018) A Markov Switching Factor-Augmented VAR Model for
  Analyzing US Business Cycles and Monetary Policy.
\newblock \emph{Oxford Bulletin of Economics and Statistics} 80(3), 575--604

\bibitem[{Huber et~al.(2018)Huber, Pfarrhofer and
  Z{\"o}rner}]{huber2018stochastic}
Huber F, Pfarrhofer M and Z{\"o}rner TO (2018) Stochastic model specification
  in Markov switching vector error correction models.
\newblock \emph{ArXiv Preprint} 1807.00529, 1--17

\bibitem[{Igan and Tan(2017)}]{igan2017capital}
Igan D and Tan Z (2017) Capital inflows, credit growth, and financial systems.
\newblock \emph{Emerging Markets Finance and Trade} 53(12), 2649--2671

\bibitem[{{IMF}(2016)}]{imf-pp-2016}
{IMF} (2016) Capital flows -- review of experience with the institutional view.
\newblock {IMF} {P}olicy {P}aper, International Monetary Fund

\bibitem[{{IMF}(2017)}]{imf-pp-2017}
{IMF} (2017) Increasing resilience to large and volatile capital flows: the
  role of macroprudential policies.
\newblock {IMF} {P}olicy {P}aper, International Monetary Fund

\bibitem[{{IMF-FSB-BIS}(2016)}]{imf-bis-pp-2016}
{IMF-FSB-BIS} (2016) Elements of effective macroprudential polivies: lessons
  from international experience.
\newblock {IMF} {P}olicy {P}aper, International Monetary Fund

\bibitem[{Kaufmann(2015)}]{kaufmann2015k}
Kaufmann S (2015) K-state switching models with time-varying transition
  distributions - Does loan growth signal stronger effects of variables on
  inflation?
\newblock \emph{Journal of Econometrics} 187(1), 82--94

\bibitem[{Kim and Nelson(1998)}]{kim1998business}
Kim CJ and Nelson CR (1998) Business cycle turning points, a new coincident
  index, and tests of duration dependence based on a dynamic factor model with
  regime switching.
\newblock \emph{Review of Economics and Statistics} 80(2), 188--201

\bibitem[{Kim et~al.(2019)Kim, Kim and Mehrotra}]{kim2019macroprudential}
Kim J, Kim S and Mehrotra A (2019) Macroprudential policy in {A}sia.
\newblock \emph{Journal of Asian Economics} 65, 101149

\bibitem[{Kim and Mehrotra(2017)}]{kim2017managing}
Kim S and Mehrotra A (2017) Managing price and financial stability objectives
  in inflation targeting economies in Asia and the Pacific.
\newblock \emph{Journal of Financial Stability} 29, 106--116

\bibitem[{Kim and Mehrotra(2018)}]{kim2018effects}
Kim S and Mehrotra A (2018) Effects of monetary and macroprudential policies --
  evidence from four inflation targeting economies.
\newblock \emph{Journal of Money, Credit and Banking} 50(5), 967--992

\bibitem[{Kim and Mehrotra(2019)}]{kim2019examining}
Kim S and Mehrotra A (2019) Examining macroprudential policy and its
  macroeconomic effects -- {S}ome new evidence.
\newblock BIS Working Papers 825, Bank for International Settlements

\bibitem[{Lepers and Mehigan(2019)}]{lepers-mehigan-2019}
Lepers E and Mehigan C (2019) The broad policy toolkit for financial stability:
  Foundations, fences, and fire doors.
\newblock OECD Working Papers on International Investment~2, Organisation for
  Economic Co-operation and Development

\bibitem[{Lepers and Mercado(2020)}]{lepers-mercado-2020}
Lepers E and Mercado R (2020) Sectoral Capital Flows: Covariates, Co-movements,
  and Controls.
\newblock SEACEN Working Paper~4, The South East Asian Central Banks Research
  and Training Centre

\bibitem[{Malsiner-Walli et~al.(2016)Malsiner-Walli, Fr{\"u}hwirth-Schnatter
  and Gr{\"u}n}]{malsinerwalli2016}
Malsiner-Walli G, Fr{\"u}hwirth-Schnatter S and Gr{\"u}n B (2016) Model-based
  clustering based on sparse finite Gaussian mixtures.
\newblock \emph{Statistics and Computing} 26(1), 303--324

\bibitem[{Meeks(2017)}]{meeks2017capital}
Meeks R (2017) Capital regulation and the macroeconomy: empirical evidence and
  macroprudential policy.
\newblock \emph{European Economic Review} 95, 125--141

\bibitem[{Meuleman and Vander~Vennet(2020)}]{meuleman2020macroprudential}
Meuleman E and Vander~Vennet R (2020) Macroprudential policy and bank systemic
  risk.
\newblock \emph{Journal of Financial Stability} 47, 100724

\bibitem[{Ostry et~al.(2012)Ostry, Ghosh, Chamon and Qureshi}]{ostry2012tools}
Ostry JD, Ghosh AR, Chamon M and Qureshi MS (2012) Tools for managing
  financial-stability risks from capital inflows.
\newblock \emph{Journal of International Economics} 88(2), 407--421

\bibitem[{Portes et~al.(2020)Portes, Beck, Buiter, Dominguez, Gros, Gross,
  Kalemli-Ozcan, Peltonen and S\'{a}nchez~Serrano}]{ESRB-2020}
Portes R, Beck T, Buiter W, Dominguez K, Gros D, Gross C, Kalemli-Ozcan S,
  Peltonen T and S\'{a}nchez~Serrano A (2020) The global dimensions of
  macroprudential policy.
\newblock Reports of the Advisory Scientific Committee of the ESRB~10, European
  Systemic Risk Board

\bibitem[{Reinhardt and Sowerbutts(2015)}]{reinhardt2015regulatory}
Reinhardt D and Sowerbutts R (2015) Regulatory arbitrage in action: evidence
  from banking flows and macroprudential policy.
\newblock Bank of England Working Paper 546, Bank of England

\bibitem[{Rey(2015)}]{rey2015dilemma}
Rey H (2015) Dilemma not trilemma: the global financial cycle and monetary
  policy independence.
\newblock Working Paper 21162, National Bureau of Economic Research

\bibitem[{Richter et~al.(2018)Richter, Schularick and Shim}]{richter-etal-2018}
Richter B, Schularick M and Shim I (2018) The macroeconomic effects of
  macroprudential policy.
\newblock BIS Working Papers 740, Bank for International Settlements

\bibitem[{Shim et~al.(2013)Shim, Bogdanova, Shek and
  Subelyte}]{shim2013database}
Shim I, Bogdanova B, Shek J and Subelyte A (2013) Database for policy actions
  on housing markets.
\newblock \emph{BIS Quarterly Review} September, 83--95

\bibitem[{Sims and Zha(1998)}]{sims1998bayesian}
Sims CA and Zha T (1998) Bayesian methods for dynamic multivariate models.
\newblock \emph{International Economic Review} , 949--968

\bibitem[{Svensson(2018)}]{svensson2018monetary}
Svensson LE (2018) Monetary policy and macroprudential policy: different and
  separate?
\newblock \emph{Canadian Journal of Economics/Revue canadienne
  d'{\'e}conomique} 51(3), 802--827

\bibitem[{Vandenbussche et~al.(2015)Vandenbussche, Vogel and
  Detragiache}]{vandenbussche2015macroprudential}
Vandenbussche J, Vogel U and Detragiache E (2015) Macroprudential policies and
  housing prices: a new database and empirical evidence for {Central, Eastern,
  and Southeastern Europe}.
\newblock \emph{Journal of Money, Credit and Banking} 47(S1), 343--377

\bibitem[{Verbeke and Lesaffre(1996)}]{verbeke1996linear}
Verbeke G and Lesaffre E (1996) A linear mixed-effects model with heterogeneity
  in the random-effects population.
\newblock \emph{Journal of the American Statistical Association} 91(433),
  217--221

\end{thebibliography}
\clearpage
\begin{appendices}\crefalias{section}{appsec}
\setcounter{equation}{0}
\renewcommand\theequation{A.\arabic{equation}}
\normalsize
\newpage

\section{Additional results}\label{app:res}

\subsection{Impulse-responses for MS specification}\label{add:MSirfs}

\begin{figure}[!htbp]
	\centering
	\includegraphics[width=0.8\linewidth]{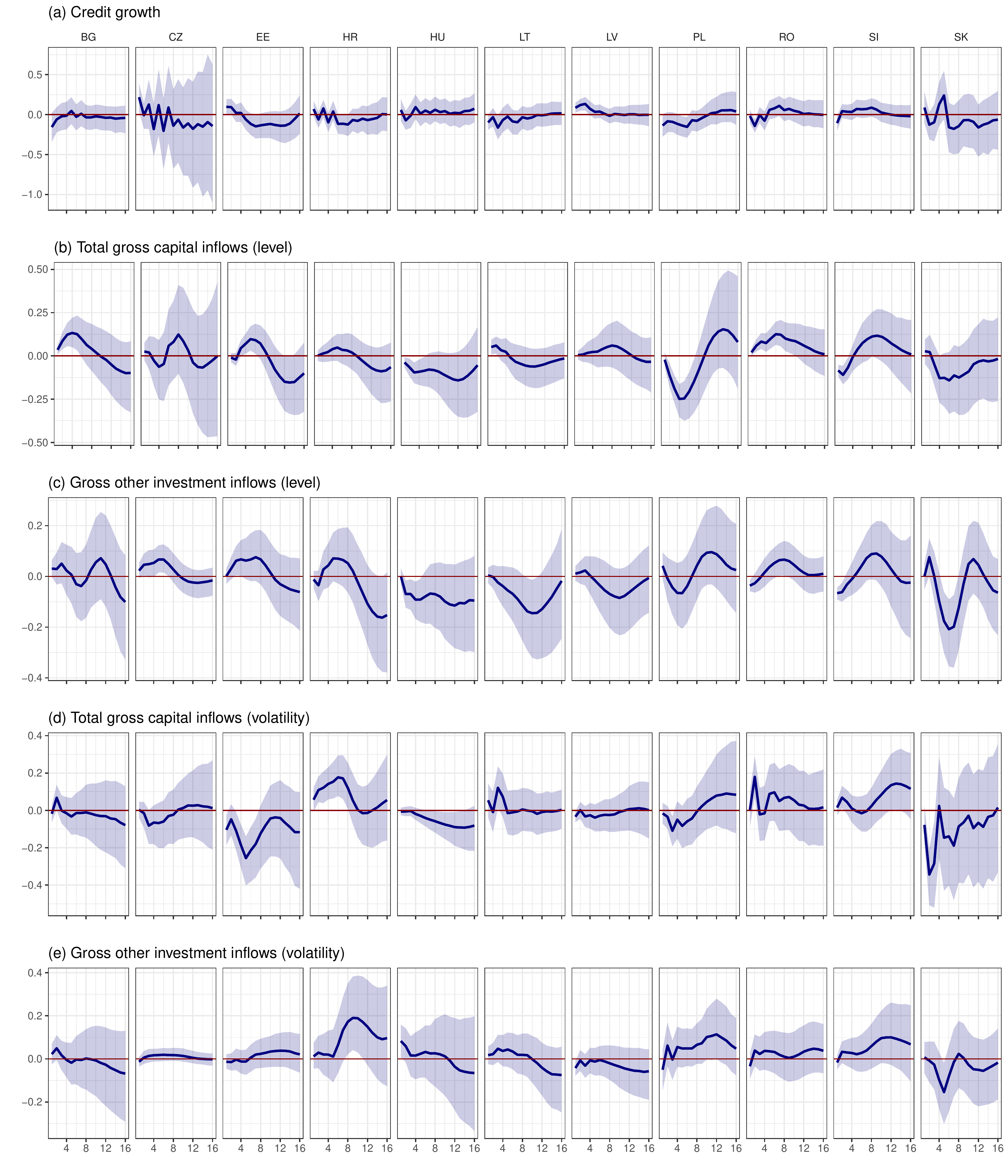}
\begin{minipage}{1\textwidth}
\footnotesize \textbf{Note:} The blue line denotes the posterior median and blue shaded areas refer to the $68$\,\% credible set.  
\end{minipage}
	\caption{High-interest rate regime (posterior median) \textbf{impulse-responses of private sector credit growth, total gross capital and gross other investment inflows} (levels \& volatilities) to a 1 SD tightening shock in the MPPI.}
	\label{fig:IRF_high_ms_level}
\end{figure}

\begin{figure}[!htbp]
	\centering
	\includegraphics[width=0.8\linewidth]{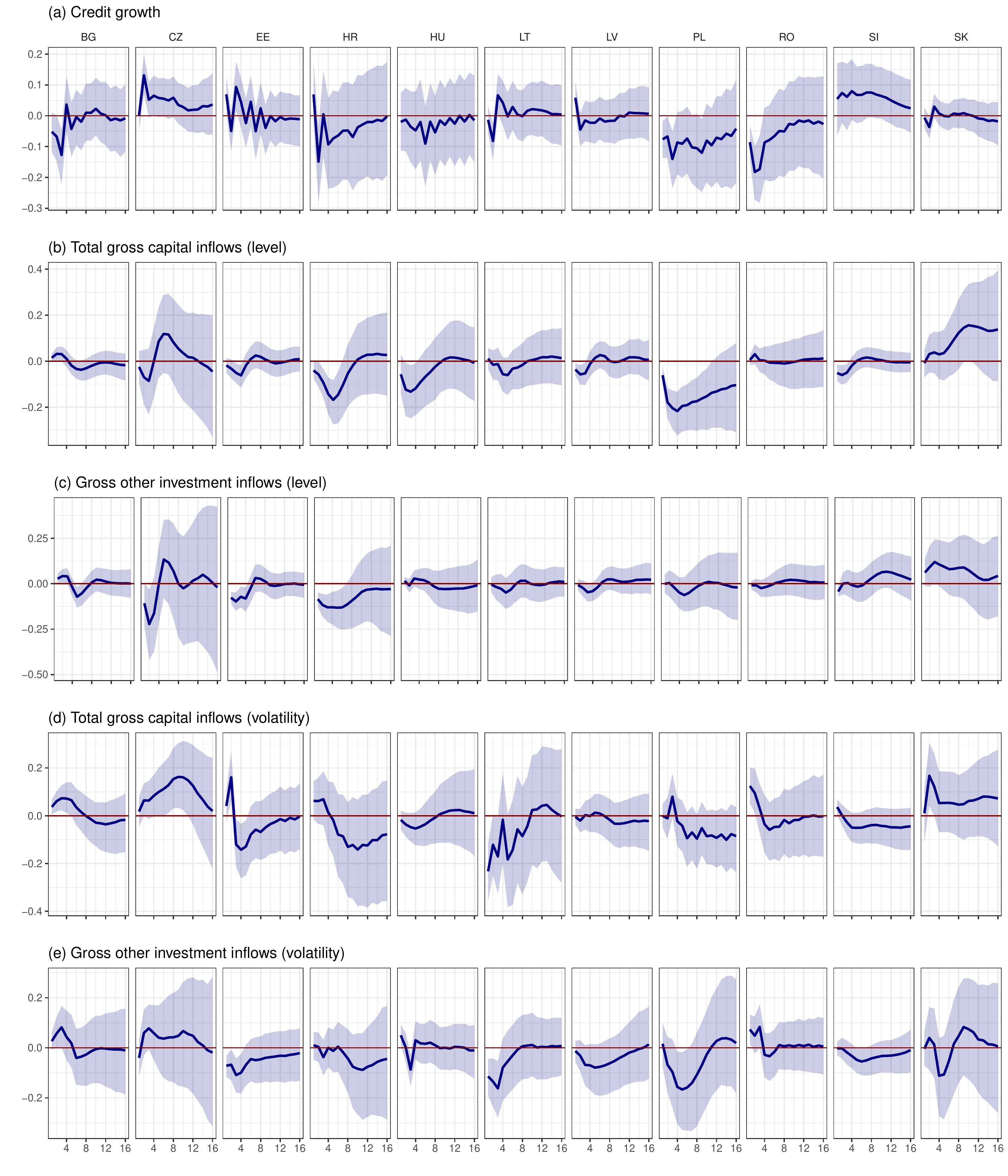}
\begin{minipage}{1\textwidth}
\footnotesize \textbf{Note:} The blue line denotes the posterior median and blue shaded areas refer to the $68$\,\% credible interval.  
\end{minipage}
	\caption{Low-interest rate regime \textbf{impulse-responses of private sector credit growth, total gross capital and gross other investment inflows} (levels \& volatilities) to a 1 SD tightening shock in the MPPI.}
	\label{fig:IRF_low_ms_level}
\end{figure}

\clearpage



\clearpage

\subsection{Deterministic regime allocation}\label{add:detregime}

\begin{figure}[!htbp]
	\centering
	\includegraphics[width=\linewidth]{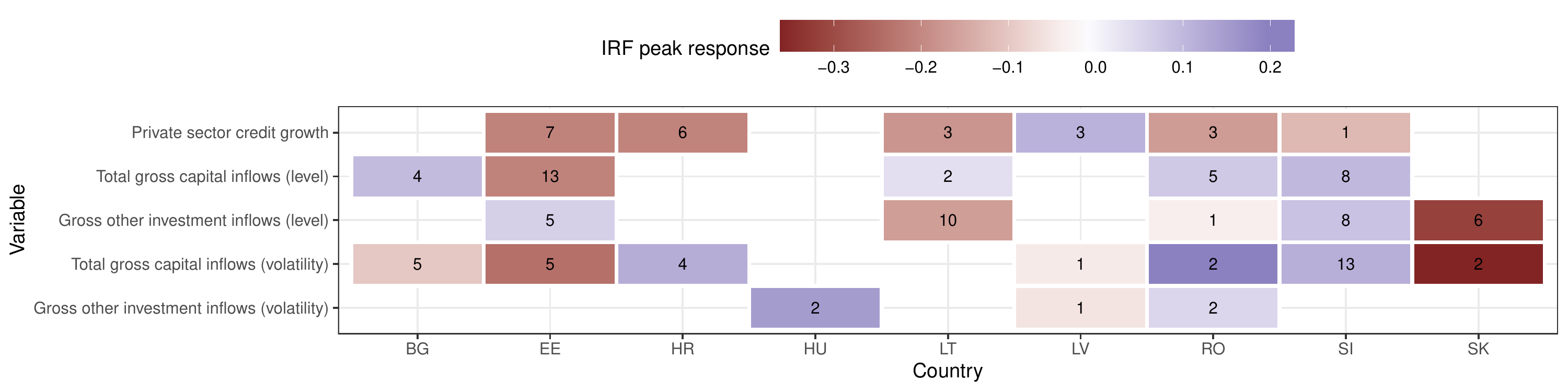}
\begin{minipage}{\textwidth}
\footnotesize \textbf{Note:} Red shaded cells denote negative and blue shaded cells denote positive responses. Cell numbers indicate the quarter after the shock at which the response reaches its peak. Empty cells refer to insignificance with respect to the $68$\,\% credible interval. For CZ and PL, we do not observe any significant responses of these variables. 
\end{minipage}
	\caption{Pre-GFC (posterior median) \textbf{peak responses of private sector credit growth, total gross capital and gross other investment inflows} (levels \& volatilities) to a 1 SD tightening shock in the MPPI.}
	\label{fig:IRF_pre_peak_mapru}
\end{figure}

\begin{figure}[!htbp]
	\centering
	\includegraphics[width=\linewidth]{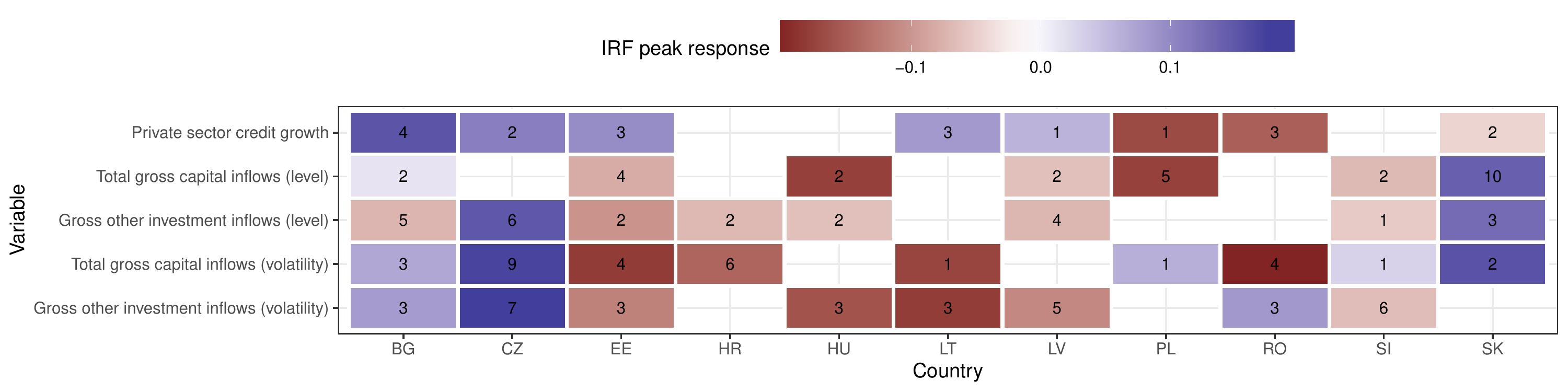}
\begin{minipage}{\textwidth}
\footnotesize \textbf{Note:} Red shaded cells denote negative and blue shaded cells denote positive responses. Cell numbers indicate the quarter after the shock at which the response reaches its peak. Empty cells refer to insignificance with respect to the $68$\,\% credible interval.  
\end{minipage}
	\caption{Post-GFC (posterior median) \textbf{peak responses of private sector credit growth, total gross capital and gross other investment inflows} (levels \& volatilities) to a 1 SD tightening shock in the MPPI.}
	\label{fig:IRF_post_peak_mapru}
\end{figure}

\clearpage

\section{Data Descriptives}
\label{app:data}

\begin{table}[!htbp]
	\caption{Variable description}
	\footnotesize{
	\centering
	 \setlength\extrarowheight{0pt}
	 \begin{tabular}{p{3.5cm}p{8cm}p{4cm}}
		\textbf{Variable} & \textbf{Description} & \textbf{Main source(s)} \\
		\hline
		\hline
        Global factor & Factor extract of equity price, credit and deposit growth for a set of 45 developed \& developing countries; no further transformation & own calculations; IMF-IFS\\
        \hline
        MPPI & Intensity-adjusted macroprudential policy index; first differences & \cite{eller-etal-taxonomy}\\
        \hline
        GDP growth & GDP volume, 2005=100, seasonally adjusted, in logarithms, quarter-on-quarter changes & IMF-IFS\\
        \hline
        Inflation rate & (Harmonized) consumer price index, 2005=100, seasonally adjusted, quarter-on-quarter change & IMF-IFS\\
        \hline
        Credit growth & Domestic banks' claims on the resident nonbank private sector, CPI deflated, seasonally adjusted, in logarithms, quarter-on-quarter changes & IMF-IFS; BIS\\
        \hline
        Short-term interest rate & Typically, three-month money market rate (per annum); no further transformation & IMF-IFS; ECB; Eurostat\\
        \hline
        Equity price growth & Equity price index, 2005=100, seasonally adjusted, in logarithms, quarter-on-quarter change & IMF-IFS; OECD\\
        \hline
        REER volatility & Real effective exchange rate, CPI-based index, seasonally adjusted, in logarithms, estimated log-variances of AR(5)-SV process & own calculations; IMF-IFS\\
        \hline
     Gross capital inflows & Cumulative four-quarter moving sums of either total capital inflows (i.e. incurrence less repayment of residents' totaled direct, portfolio and other investment (OI) liabilities vis-\'{a}-vis nonresidents) or OI inflows only (BPM6 definition), as percentage of nominal GDP & IMF-IFS\\
        \hline
     Gross capital outflows & Cumulative four-quarter moving sums of either total capital outflows (i.e. residents' acquisition less disposal of totaled direct, portfolio and other investment (OI) assets abroad) or OI outflows only (BPM6 definition), as percentage of nominal GDP & IMF-IFS\\
        \hline
        Capital flow volatilities \newline \scriptsize{of in- and outflows} & Estimated log-variances of AR(5)-SV process on respective capital flow series & own calculations; IMF-IFS\\
        \hline
	\end{tabular}
	}
	
	\vspace{1ex}
	
	{\raggedright \footnotesize \textbf{Notes:} This table presents the variables included in the country-specific FAVAR models, a short description and their corresponding transformations for estimation as well as sources from where they were gathered. Seasonal adjustment was conducted using the Census X12 method. A few capital flow 
	 series were not satisfactorily available at quarterly frequency at the beginning of the sample; we used the corresponding annual figures and the quarterly dynamics of the remaining sample for data interpolation. If for equity prices data from the IMF-IFS was missing, dynamics from OECD series on equity prices or from GDP growth were used for interpolation. Moreover, in a few cases where the short-term interest rate was missing, we used the dynamics of the deposit rate for data interpolation similarly to \cite{eller-huber-schuberth-2020}. In cases where a few observations were missing at the beginning or the end of the sample, we used the average of the subsequent or previous four quarters to fill these gaps. All variables were standardized prior to estimation by subtracting the mean and dividing them by their standard deviations. \par}
\label{tab:data}
\end{table}


%

\end{appendices}

\end{document}